\documentclass[aps,prc,twocolumn,showpacs,floatfix,nofootinbib,preprintnumbers,superscriptaddress,amsmath,amssymb]{revtex4-1}

\usepackage{graphicx}
\usepackage{epsfig}
\usepackage{bm}
\usepackage{color}
\usepackage{float}
\usepackage{dcolumn}
\usepackage{multirow} 
\usepackage{hyperref}

\graphicspath{{.}}

\newcommand{\beq}{\begin{equation}}
\newcommand{\eeq}{\end{equation}}
\newcommand{\beqa}{\begin{eqnarray}}
\newcommand{\eeqa}{\end{eqnarray}}

\newcommand{\braket}[2]{\left\langle #1 \vert #2 \right\rangle}

\begin{document}

\title{Coupled-cluster calculations of nucleonic matter}

\author{G.~Hagen} \affiliation{Physics Division, Oak Ridge National
  Laboratory, Oak Ridge, TN 37831, USA} 
  \affiliation{Department of
  Physics and Astronomy, University of Tennessee, Knoxville, TN 37996,
  USA} 

\author{T.~Papenbrock} \affiliation{Department
  of Physics and Astronomy, University of Tennessee, Knoxville, TN
  37996, USA} \affiliation{Physics Division, Oak Ridge National
  Laboratory, Oak Ridge, TN 37831, USA} 

\author{A.~Ekstr\"om} \affiliation{Department of Physics and Center of
  Mathematics for Applications, University of Oslo, N-0316 Oslo,
  Norway} \affiliation{National Superconducting Cyclotron Laboratory,
  Michigan State University, East Lansing, MI 48824, USA}

\author{K.~A.~Wendt} \affiliation{Department
  of Physics and Astronomy, University of Tennessee, Knoxville, TN
  37996, USA} \affiliation{Physics Division, Oak Ridge National
  Laboratory, Oak Ridge, TN 37831, USA} 

\author{G.~Baardsen} \affiliation{Department of Physics and Center of
  Mathematics for Applications, University of Oslo, N-0316 Oslo,
  Norway}

\author{S.~Gandolfi} \affiliation{Theoretical Division, 
Los Alamos National Laboratory Los Alamos, NM 87545}

\author{M.~Hjorth-Jensen} \affiliation{Department of Physics
  and Center of Mathematics for Applications, University of Oslo,
  N-0316 Oslo, Norway} \affiliation{National Superconducting Cyclotron
  Laboratory, Michigan State University, East Lansing, MI 48824,
  USA} \affiliation{Department of Physics and Astronomy, Michigan
  State University, East Lansing, MI 48824, USA} 

\author{C.~J.~Horowitz} \affiliation{Indiana University, Bloomington, IN 47405, USA}

\begin{abstract}
  \noindent {\bf Background:} The equation of state (EoS) of nucleonic
  matter is central for the understanding of bulk nuclear properties,
  the physics of neutron star crusts, and the energy release in
  supernova explosions. Because nuclear matter exhibits a finely tuned
  saturation point, its EoS also constrains nuclear interactions. \\
  {\bf Purpose:} This work presents coupled-cluster calculations of
  infinite nucleonic matter using modern interactions from chiral
  effective field theory (EFT). It assesses the role of correlations
  beyond particle-particle and hole-hole ladders, and the role of
  three-nucleon-forces (3NFs)
  in nuclear matter calculations with chiral interactions.\\
  {\bf Methods:} This work employs the optimized nucleon-nucleon
  ($NN$) potential NNLO$_{\rm opt}$ at next-to-next-to leading-order,
  and presents coupled-cluster computations of the EoS for symmetric
  nuclear matter and neutron matter. The coupled-cluster method
  employs up to selected triples clusters and the single-particle
  space consists of a momentum-space lattice. We compare our results
  with benchmark calculations and control finite-size effects and
  shell oscillations via twist-averaged
  boundary conditions. \\
  {\bf Results:} We provide several benchmarks to validate the
  formalism and show that our results exhibit a good convergence
  toward the thermodynamic limit.  Our calculations agree well with
  recent coupled-cluster results based on a partial wave expansion and
  particle-particle and hole-hole ladders. For neutron matter at low
  densities, and for simple potential models, our calculations agree
  with results from quantum Monte Carlo computations. While neutron
  matter with interactions from chiral EFT is perturbative, symmetric
  nuclear matter requires nonperturbative approaches. Correlations
  beyond the standard particle-particle ladder approximation yield
  non-negligible contributions. The saturation point of symmetric
  nuclear matter is sensitive to the employed 3NFs and the employed
  regularization scheme. 3NFs with nonlocal cutoffs exhibit a
  considerably improved convergence than their local cousins.  We are
  unable to find values for the parameters of the short-range part of
  the local 3NF that simultaneously yield acceptable values for the
  saturation point in symmetric nuclear matter and the binding
  energies of light nuclei. \\
  {\bf Conclusions:} Coupled-cluster calculations with nuclear
  interactions from chiral EFT yield nonperturbative results for the
  EoS of nucleonic matter. Finite-size effects and effects of
  truncations can be controlled. For the optimization of chiral
  forces, it might be useful to include the saturation point of
  symmetric nuclear matter.
\end{abstract}


\maketitle

\section{Introduction}\label{sec:intro}
Bulk nucleonic matter is interesting for several reasons. The EoS of
neutron matter, for instance, determines properties of supernova
explosions~\cite{burrows2013}, and of neutron
stars~\cite{weber1999,hh2000,lattimer2007,sammarruca2010,lattimer2012,hebeler2012f},
and it links the latter to neutron radii in atomic
nuclei~\cite{brown2000,horowitz2001,gandolfi2012} and symmetry
energy~\cite{tsang2012,steiner2012}. Likewise, the compressibility of
nuclear matter is probed in giant dipole
excitations~\cite{shlomo1993}, and the symmetry energy of nuclear
matter is related to the difference between proton and neutron radii
in atomic nuclei~\cite{abrahamyan2012,reinhard2013,erler2013}. The
saturation point of nuclear matter determines bulk properties of
atomic nuclei, and is therefore an important constraint for nuclear
energy-density functionals and mass models (see,
e.g., Refs.~\cite{kortelainen2010,lunney2003}).

The determination and our understanding of the EoS for nuclear matter
is intimately linked with our capability to solve the nuclear
many-body problem. Here, correlations beyond the mean field play an
important role.  Theoretical studies of nuclear matter and the
pertinent EoS span back to the very early days of nuclear many-body
physics. Early computations are nicely described in the 1967 review by
Day \cite{day1967}. These early calculations were performed using
Brueckner-Bethe-Goldstone theory \cite{brueckner1954,brueckner1955},
see Refs.\cite{hh2000,baldo2012,baldo2012a} for recent reviews and
developments.  In these calculations, mainly particle-particle
correlations were summed to infinite order.  Other correlations were
often included in a perturbative way. Coupled-cluster calculations of
nuclear matter were performed already during the late 1970s and early
1980s~\cite{kummel1978,day1981}. In recent years, there has been a
considerable algorithmic development of first-principle methods for
solving the nuclear many-body problem. A systematic inclusion of other
correlations in a non-perturbative way are nowadays accounted for in
Monte Carlo methods
\cite{carlson2003,gandolfi2009,gezerlis2010,lovato2012,gezerlis2013},
self-consistent Green's function approaches
\cite{dickhoff2004,soma2008, soma2012, baldo2012a,carbone2013} and
nuclear density functional theory \cite{lunney2003,erler2013}.

Similar progress has been made in the derivation of nuclear forces
based on chiral EFT~\cite{machleidt2011,epelbaum2009}. Nuclear
Hamiltonians from chiral EFT are now used routinely in nucleonic
matter calculations, with the 3NFs~\cite{vankolck1994,epelbaum2002}
being front and center of many
studies~\cite{soma2008,epelbaum2009b,hebeler2010b,hebeler2011,holtjw2012,
hammer2013,krueger2013,carbone2013,coraggio2013}. We note finally that
there are also approaches to nucleonic matter based on lattice quantum
chromodynamics~\cite{tetsuo2013}.

In this work we study the EoS of nucleonic matter, using modern $NN$
interactions and 3NFs from chiral EFT, and an implementation of the
coupled-cluster method~\cite{dean2004,kowalski2004} that has become a
standard in quantum
chemistry~\cite{bartlett2007,shavittbartlett2009}. We employ a
Cartesian momentum space basis with periodic boundary conditions,
similar to the recent coupled-cluster based calculations of the
electron gas \cite{shepherd2012,roggero2013}. Our calculations are
based on coupled cluster with doubles (CCD) approximation
\cite{harris1992,freeman1977,bishop1978}. This is the lowest-order
truncation for closed-shell systems in a momentum-space basis, and we
will also explore the role of selected triples clusters. We employ a
recent parameterization \cite{ekstrom2013} of the $NN$ force from
chiral EFT at next-to-next-to-leading order, with inclusion of the 3NF
that enters at the same chiral order.

This paper is organized as follows. In the next Section we present the
coupled-cluster formalism for infinite matter that includes 3NFs and
perturbative triples corrections. The calculations are performed in
Cartesian coordinates with a discrete momentum basis and twisted
periodic boundary conditions~\cite{gros1992,gros1996,lin2001}. This
avoids the tedious partial-wave expansion of the nuclear forces, and
it eases considerably the numerical evaluation of 3NFs.  Averaging
over twisted periodic boundary conditions minimizes finite-size
effects and provides us with a good convergence towards the
thermodynamic limit. Section~\ref{sec:method} also presents
computational results for finite-size effects and a few benchmark
calculations.  Our results for symmetric nuclear matter and pure
neutron matter are presented in Sect.~\ref{sec:results}. Concluding
remarks are given in Sect.~\ref{sec:summary}.

\section{Method}\label{sec:method}
In this Section we present the coupled-cluster formalism for infinite
matter. We discuss the inclusion and treatment of $NN$ forces and 3NFs
from chiral effective field theory (EFT), correlations up to
three-particle-three-hole excitations and finite-size effects. Several
benchmark calculations give us confidence in the validity of our
approach.  \subsection{Interaction and model space}
Our Hamiltonian is 
\[
H = T_{\rm kin}+V_{NN} + V_{\rm 3NF} \ . 
\]
Here, $T_{\rm kin}$ denotes the kinetic energy, and $V_{NN}$ and $V_{\rm 3NF}$
denote the translationally invariant $NN$ interaction and 3NF.  The
$NN$ interaction and 3NF are from chiral effective field
theory~\cite{epelbaum2009,machleidt2011} at next-to-next-to-leading
order (NNLO). We employ the parameterization NNLO$_{\rm opt}$ for the
$NN$ interaction~\cite{ekstrom2013}, and the local
3NF~\cite{navratil2007}. This 3NF has a local regulator, i.e. the
cutoff is in the momentum transfer, and thereby differs from
implementations of the 3NF~\cite{epelbaum2002} that employ the cutoff
in the relative Jacobi momenta.  We note that the numerical
implementation of the 3NF in the discrete momentum basis is much
simpler than in the harmonic oscillator basis commonly used for finite
nuclei, because essentially no transformation of matrix elements is
necessary. Nevertheless, the sheer number of matrix elements (and
associated function calls) of the 3NF is huge, and this is
computationally still a limiting factor.

For the model space, we choose a cubic lattice in momentum space with
$(2n_{\max}+1)^3$ momentum points. The spin (spin-isospin) degeneracy
of each momentum point is $g_s=2$ ($g_s=4$) for pure neutron matter
(nuclear matter). Thus, filling of the lattice yields shell closures
for ``Fermi spheres'' with $g_s n$ fermions, and $n=1, 7, 19, 27, 33,
57, \ldots$. We note that one could also use non-cubic lattices. Any
periodic lattice permits one to implement momentum conservation
exactly. For fixed particle number $A=g_s n$ and density $\rho=g_s
k_F^3/(6\pi^2)$ (or Fermi momentum $k_F$), one computes the volume of
the cubic box $V=L^3=A/\rho$, and the box length $L$ that determines
the lattice spacing $\Delta k=2\pi/L$.  We note that the computed
results exhibit a dependence on the shell closure $n$. However,
Subsection~\ref{finite} shows that shell effects and finite-size
effects can be mitigated and controlled, and that the dependence on
the parameter $n$ becomes very small.

The second parameter of our lattice is $n_{\max}$. We note that
$n_{\max} \Delta k$ is the momentum cutoff of our single-particle
basis. One has to increase $n_{\max}$ until the computed results (e.g.
the energy per nucleon) is practically independent of this parameter.
For the results reported below we find that $n_{\max}=4$ is
sufficient.

\subsection{Coupled-cluster theory for infinite systems}
In this Section we present the coupled-cluster equations for nucleonic
matter. Our calculations of nucleonic matter are based on the recently
optimized chiral nucleon-nucleon interaction at NNLO$_{\mathrm opt}$
\cite{ekstrom2013} with the 3NF at the same chiral order. The
low-energy constants (LECs) of the 3NF were determined by fitting the
constants $c_E$ and $c_D$ to reproduce the experimental half-life and
binding energy of the triton.  The optimized 3NF LECs are $c_E =-0.389
$ and $c_D =-0.39 $. With these values the $^4$He binding energy is
$-28.47$~MeV \cite{navratil_gazit2013}. We employ single-particle
states
\begin{equation} 
\nonumber
\vert \vec p, s_z, t_z\rangle \equiv \vert \mathbf{k}\rangle, 
\end{equation}
with momentum $\vec{k}$, spin projection $s_z$ and isospin projection
$t_z$. Discrete values of the momentum variable $ \vec p = \hbar \vec k$ result
from periodic boundary conditions in a cubic box with length $L$, that is 
\[
k_{n_i} = \frac{2\pi n_i}{L}, \:\: n_i = 0, \pm 1, \ldots \pm n_{\mathrm{max}}, \:\: i = x,y,z.
\]
In this basis, the nuclear Hamiltonian with nucleon-nucleon and three-nucleon
interactions is 
\begin{eqnarray}\label{eq:ham}
H  & = & \sum_{pq}\braket{\mathbf{k}_{p}}{T_{\rm kin}|\mathbf{k}_{q}}a^\dagger_{p} a_q \nonumber\\&+& 
\frac{1}{4}\sum_{pqrs}\braket{\mathbf{k}_{p}\mathbf{k}_{q}}{V_{\mathrm{NN}}|\mathbf{k}_{r}\mathbf{k}_{s}}
a^\dagger_{p} a^\dagger_q a_{s}a_r \\ 
& + &  
\frac{1}{36}\sum_{pqrstu}\braket{\mathbf{k}_{p}\mathbf{k}_{q}\mathbf{k}_{r}}{V_{\mathrm{3NF}}|\mathbf{k}_{s}\mathbf{k}_{t}\mathbf{k}_{u}}
a^\dagger_{p} a^\dagger_q a^\dagger_{r}a_ua_ta_s.\nonumber  
\end{eqnarray}
The kinetic energy is diagonal in the
discrete momentum basis $\braket{\mathbf{k}_{p}}{T_{\rm kin}|\mathbf{k}_{q}} =
\frac{\hbar^2}{2m}\vec{k}_p^2\delta_{pq}$. The operators $a_p^\dagger$ and $a_p$ create and annihilate a nucleon in state $|{\bf k}_p\rangle$, respectively. 

The discrete momentum basis allows us to respect translational
invariance of the $NN$ potential and the 3NFs. 
Momentum is conserved, meaning that the two- and three-body matrix
elements of the Hamiltonian~(\ref{eq:ham}) vanish unless
\[
  \vec{k}_{p}+ \vec{k}_{q}  =   \vec{k}_{r}+\vec{k}_{s} ,
\]
and 
\[
  \vec{k}_{p}+\vec{k}_{q}+\vec{k}_{r}  =  \vec{k}_{s}+\vec{k}_{t}+\vec{k}_{u} . 
\]
Note also that the chiral nucleon-nucleon and three-nucleon
interactions conserve the total isospin projection, but not the total
spin projection.

In single-reference coupled-cluster theory the correlated 
wave-function is written in the form 
\[
  \vert \Psi\rangle = e^T\vert\Phi_0\rangle . 
\]
Here $\vert \Phi_0\rangle = \prod_{i=1}^A a^\dagger_i|0\rangle$ is a
product state and serves as the reference. The cluster operator $T$ is
a linear combination of $n$-particle-$n$-hole ($np$-$nh$) excitation
operators, i.e. $T=T_1+T_2+\ldots+T_n$. In the discretized momentum
basis the reference state is the closed shell Fermi vacuum, and is
obtained by filling the $A$ states with the lowest kinetic energy. We
limit ourselves to spin saturated reference state, meaning that  each momentum
orbital of the reference state is doubly occupied. In this case the
nuclear interaction does not induce $1p$-$1h$ excitations of the
reference state, and we have $T_1 = 0$. Thus, the cluster operator
becomes
\[
T = \frac{1}{4}\sum_{ijab}  \braket{\mathbf{k}_{a}\mathbf{k}_{b}}{t|\mathbf{k}_{i}\mathbf{k}_{j}} a^\dagger_{a} a^\dagger_b a_{j}a_i + \ldots .
\]
Here and in what follows, indices $i,j,k,l$ ($a,b,c,d$) label occupied
(unoccupied) states.  Truncating $T$ at the 2$p$-2$h$ excitation level
($T\approx T_2$) gives the coupled-cluster doubles (CCD)
approximation. The CCD energy and amplitude equations can be written
in compact form
\begin{eqnarray} 
  \label{eq:ccd_energy}
  E_{\mathrm{CCD}} &  = &  E_0+\langle \Phi_0 \vert \overline{H}_N \vert \Phi_0 \rangle, \\ 
  \label{eq:ccd}
  0 & = &  \langle \Phi_{ij}^{ab} \vert \overline{H}_N \vert \Phi_0 \rangle.
\end{eqnarray}
Here
\beqa
\label{evac}
E_0 &=& \langle \Phi_0 \vert H \vert \Phi_0\rangle \nonumber\\
&=& \sum_i \braket{\mathbf{k}_{i}}{f|\mathbf{k}_{i}} 
+\frac{1}{2}\sum_{i,j}
\braket{\mathbf{k}_{i}\mathbf{k}_{j}}{v|\mathbf{k}_{i}\mathbf{k}_{j}}\nonumber\\
&&+\frac{1}{6}\sum_{ijk}\braket{\mathbf{k}_{i}\mathbf{k}_{j}\mathbf{k}_{l}}{w|\mathbf{k}_{i}\mathbf{k}_{j}\mathbf{k}_{l}}
\eeqa
is the vacuum
expectation value (which in the case of no 1$p$-1$h$ corresponds to
the Hartree-Fock energy), $ \vert \Phi_{ij}^{ab}\rangle $ is a
2$p$-2$h$ excitation of the reference state, and $ \overline{H}_N
\equiv e^{-T}H_Ne^T$ is the similarity transformation of the 
normal-ordered Hamiltonian
\begin{eqnarray}
\nonumber
H_{N}  & = & \sum_{pq}\braket{\mathbf{k}_{p}}{f|\mathbf{k}_{q}}:a^\dagger_{p} a_q: \nonumber \\ 
& + &  
\frac{1}{4}\sum_{pqrs}\braket{\mathbf{k}_{p}\mathbf{k}_{q}}{v|\mathbf{k}_{r}\mathbf{k}_{s}}
:a^\dagger_{p} a^\dagger_q a_{s}a_r: \\ 
& + &  
\nonumber 
\frac{1}{36}\sum_{pqrstu}\braket{\mathbf{k}_{p}\mathbf{k}_{q}\mathbf{k}_{r}}{w|\mathbf{k}_{s}\mathbf{k}_{t}\mathbf{k}_{u}}
:a^\dagger_{p} a^\dagger_q a^\dagger_{r}a_ua_ta_s:.  
\label{eq:ham_N}
\end{eqnarray}
Here $:a^\dagger_{p} \ldots a_{p'}\ldots: $ is the normal ordered
string of operators with respect to the reference state. The
normal-ordered one-body operator is given in terms of the Fock matrix
elements
\begin{eqnarray}
  \label{fock}
  \nonumber
  \braket{\mathbf{k}_{p}}{f|\mathbf{k}_{q}} &  = & 
  \braket{\mathbf{k}_{p}}{t|\mathbf{k}_{q}}+\sum_{i}\braket{\mathbf{k}_{p}\mathbf{k}_{i}}{V_{\mathrm NN}|\mathbf{k}_{q}\mathbf{k}_{i}} \\ 
  & + &  
  \frac{1}{2}\sum_{ij} \braket{\mathbf{k}_{p}\mathbf{k}_{i}\mathbf{k}_{j}}{V_{\mathrm{3NF}}|\mathbf{k}_{q}\mathbf{k}_{i}\mathbf{k}_{j}}.
\end{eqnarray}

The normal-ordered two-body operator has matrix elements
\begin{eqnarray}\label{no2}
  \braket{\mathbf{k}_{p}\mathbf{k}_{q}}{v|\mathbf{k}_{r}\mathbf{k}_{s}} & = &  
  \braket{\mathbf{k}_{p}\mathbf{k}_{q}}{V_{\mathrm NN}|\mathbf{k}_{r}\mathbf{k}_{s}} \nonumber \\ 
  & + & 
  \sum_i \braket{\mathbf{k}_{p}\mathbf{k}_{q}\mathbf{k}_{i}}{V_{\mathrm{3NF}}|\mathbf{k}_{i}\mathbf{k}_{r}\mathbf{k}_{s}}.
\end{eqnarray}

Finally, the normal-ordered three-body operator $w$ has matrix elements
\begin{equation}
\label{no3}
  \braket{\mathbf{k}_{p}\mathbf{k}_{q}\mathbf{k}_{r}}{w|\mathbf{k}_{s}\mathbf{k}_{t}\mathbf{k}_{u}} = 
  \braket{\mathbf{k}_{p}\mathbf{k}_{q}\mathbf{k}_{r}}{V_{\mathrm{3NF}}|\mathbf{k}_{s}\mathbf{k}_{t}\mathbf{k}_{u}}.
\end{equation}
 
In most of this work, we will neglect all elements of $w$ when solving
the CCD equations. In this normal-ordered two-body approximation, the
3NF enters in the vacuum expectation value~(\ref{evac}), the Fock
matrix~(\ref{fock}), and the normal-ordered two-body
operator~(\ref{no2}), but the three-body operator $w$ that changes the
orbitals of all three nucleons is neglected.

We note that coupled-cluster theory with full inclusion of 3NFs was
worked out in the singles and doubles approximation (CCSD)
~\cite{hagen2007a}, and very recently with triples corrections
included~\cite{binder2013}. 

For an efficient numerical implementation one writes the CCD equations
(\ref{eq:ccd}) in a factorized (quasi-linear) form,
\begin{eqnarray} \label{eq:t2amp}
  0 &=&  \braket{\mathbf{k}_{a}\mathbf{k}_{b}}{v|\mathbf{k}_{i}\mathbf{k}_{j}} \nonumber \\
  & + &  P(ab) \sum_c \braket{\mathbf{k}_{b}}{\chi|\mathbf{k}_{c}} 
  \braket{\mathbf{k}_{a}\mathbf{k}_{c}}{t|\mathbf{k}_{i}\mathbf{k}_{j}} \nonumber \\
  & - &  P(ij)\sum_k  \braket{\mathbf{k}_{k}}{\chi|\mathbf{k}_{j}} 
  \braket{\mathbf{k}_{a}\mathbf{k}_{b}}{t|\mathbf{k}_{i}\mathbf{k}_{k}} \nonumber \\
  & + & \frac{1}{2}\sum_{cd}\braket{\mathbf{k}_{a}\mathbf{k}_{b}}{\chi|\mathbf{k}_{c}\mathbf{k}_{d}}
  \braket{\mathbf{k}_{c}\mathbf{k}_{d}}{t|\mathbf{k}_{i}\mathbf{k}_{j}} \nonumber \\
  & + & \frac{1}{2}\sum_{kl}\braket{\mathbf{k}_{a}\mathbf{k}_{b}}{t|\mathbf{k}_{k}\mathbf{k}_{l}}
  \braket{\mathbf{k}_{k}\mathbf{k}_{l}}{\chi|\mathbf{k}_{i}\mathbf{k}_{j}} \nonumber \\
  & + & P(ij)P(ab)\sum_{kc}\braket{\mathbf{k}_{a}\mathbf{k}_{c}}{t|\mathbf{k}_{i}\mathbf{k}_{k}} 
  \braket{\mathbf{k}_{k}\mathbf{k}_{b}}{\chi|\mathbf{k}_{c}\mathbf{k}_{j}} \nonumber \\
  & + & \frac{1}{2}P(ij)\sum_{cdk}\braket{\mathbf{k}_{a}\mathbf{k}_{k}\mathbf{k}_{b}}{w|
    \mathbf{k}_{i}\mathbf{k}_{c}\mathbf{k}_{d}}\braket{\mathbf{k}_{c}\mathbf{k}_{d}}{t|\mathbf{k}_{k}\mathbf{k}_{j}} 
  \nonumber \\ 
  & - & \frac{1}{2}P(ab)\sum_{ckl}\braket{\mathbf{k}_{a}\mathbf{k}_{k}\mathbf{k}_{l}}{w|
    \mathbf{k}_{i}\mathbf{k}_{c}\mathbf{k}_{j}}\braket{\mathbf{k}_{c}\mathbf{k}_{b}}{t|\mathbf{k}_{k}\mathbf{k}_{l}} \ .
\end{eqnarray}
Here, $P(pq)  \equiv  1 - P_{pq}$ is an antisymmetrization operator,
and we employed the intermediates
\begin{eqnarray} 
\label{eq:t2fac}
  \nonumber
  \lefteqn{\braket{\mathbf{k}_{b}}{\chi|\mathbf{k}_{c}} =\braket{\mathbf{k}_{b}}{f|\mathbf{k}_{c}}} \\
  \nonumber 
  & - &    
  \frac{1}{2}\sum_{kld}\braket{\mathbf{k}_{b}\mathbf{k}_{d}}{t|\mathbf{k}_{k}\mathbf{k}_{l}} 
  \braket{\mathbf{k}_{k}\mathbf{k}_{l}}{v|\mathbf{k}_{c}\mathbf{k}_{d}} \\
  &+& {1\over 4} \sum_{edkl} \braket{\mathbf{k}_{k}\mathbf{k}_{l}\mathbf{k}_{b}}{w|\mathbf{k}_{e}\mathbf{k}_{d}\mathbf{k}_{c}} 
  \braket{\mathbf{k}_{e}\mathbf{k}_{d}}{t|\mathbf{k}_{k}\mathbf{k}_{l}}, 
\end{eqnarray}
\begin{eqnarray}
\label{eq:t2fac2} 
  \nonumber
  \lefteqn{\braket{\mathbf{k}_{k}}{\chi|\mathbf{k}_{j}}  =   \braket{\mathbf{k}_{k}}{f|\mathbf{k}_{j}}} \\
  \nonumber 
  & + &  
  \frac{1}{2}\sum_{kcd} \braket{\mathbf{k}_{k}\mathbf{k}_{l}}{v|\mathbf{k}_{c}\mathbf{k}_{d}} 
  \braket{\mathbf{k}_{c}\mathbf{k}_{d}}{t|\mathbf{k}_{j}\mathbf{k}_{l}}  \\ 
  &+& {1\over 4} \sum_{cdln} \braket{\mathbf{k}_{l}\mathbf{k}_{n}\mathbf{k}_{k}}{w|\mathbf{k}_{c}\mathbf{k}_{d}\mathbf{k}_{j}} 
  \braket{\mathbf{k}_{c}\mathbf{k}_{d}}{t|\mathbf{k}_{l}\mathbf{k}_{n}}, 
\end{eqnarray}
\begin{eqnarray} 
\label{eq:t2fac3}
  \nonumber
  \lefteqn{ \braket{\mathbf{k}_{k}\mathbf{k}_{l}}{\chi|\mathbf{k}_{i}\mathbf{k}_{j}}  =  
  \braket{\mathbf{k}_{k}\mathbf{k}_{l}}{v|\mathbf{k}_{i}\mathbf{k}_{j}}} \\ 
  \nonumber
  & + & \frac{1}{2}\sum_{cd}\braket{\mathbf{k}_{k}\mathbf{k}_{l}}{v|\mathbf{k}_{c}\mathbf{k}_{d}}
  \braket{\mathbf{k}_{c}\mathbf{k}_{d}}{t|\mathbf{k}_{i}\mathbf{k}_{j}} \\ 
  & + &  
  \frac{1}{2}P(ij)\sum_{cdn}\braket{\mathbf{k}_{n}\mathbf{k}_{k}\mathbf{k}_{l}}{w|\mathbf{k}_{c}\mathbf{k}_{d}\mathbf{k}_{j}} 
  \braket{\mathbf{k}_{c}\mathbf{k}_{d}}{t|\mathbf{k}_{n}\mathbf{k}_{i}}, 
\end{eqnarray} 
\begin{eqnarray} 
\label{eq:t2fac4}
  \nonumber
  \lefteqn{ \braket{\mathbf{k}_{k}\mathbf{k}_{b}}{\chi|\mathbf{k}_{c}\mathbf{k}_{j}} 
   = 
  \braket{\mathbf{k}_{k}\mathbf{k}_{b}}{v|\mathbf{k}_{c}\mathbf{k}_{j}}} \\ 
  \nonumber
  & + &  
  \sum_{ld}\braket{\mathbf{k}_{k}\mathbf{k}_{l}}{v|\mathbf{k}_{c}\mathbf{k}_{d}} 
  \braket{\mathbf{k}_{c}\mathbf{k}_{d}}{t|\mathbf{k}_{l}\mathbf{k}_{j}} \\
  \nonumber
  & - &  
  \frac{1}{2}\sum_{dln}\braket{\mathbf{k}_{l}\mathbf{k}_{k}\mathbf{k}_{n}}{w|\mathbf{k}_{d}\mathbf{k}_{j}\mathbf{k}_{c}} 
  \braket{\mathbf{k}_{d}\mathbf{k}_{b}}{t|\mathbf{k}_{l}\mathbf{k}_{n}} \\ 
  & + &  
  \frac{1}{2}\sum_{del}\braket{\mathbf{k}_{l}\mathbf{k}_{k}\mathbf{k}_{b}}{w|\mathbf{k}_{d}\mathbf{k}_{e}\mathbf{k}_{c}} 
  \braket{\mathbf{k}_{d}\mathbf{k}_{e}}{t|\mathbf{k}_{l}\mathbf{k}_{j}}, 
\end{eqnarray}
\begin{eqnarray}
\label{eq:t2fac5} 
 \lefteqn{\braket{\mathbf{k}_{a}\mathbf{k}_{b}}{\chi|\mathbf{k}_{c}\mathbf{k}_{d}} 
   = 
  \braket{\mathbf{k}_{a}\mathbf{k}_{b}}{v|\mathbf{k}_{c}\mathbf{k}_{d}}} \nonumber\\ 
& - &  
  \frac{1}{2}P(ab)\sum_{ekl}\braket{\mathbf{k}_{k}\mathbf{k}_{l}\mathbf{k}_{b}}{w|\mathbf{k}_{e}\mathbf{k}_{c}\mathbf{k}_{d}} 
  \braket{\mathbf{k}_{e}\mathbf{k}_{a}}{t|\mathbf{k}_{k}\mathbf{k}_{l}}.
\end{eqnarray}

In Eqs.~(\ref{eq:t2amp}, \ref{eq:t2fac}, \ref{eq:t2fac2},
\ref{eq:t2fac3}, \ref{eq:t2fac4}) and (\ref{eq:t2fac5}) the
numerically expensive sums that involve products of two-body operators
can all be implemented efficiently as matrix-matrix
multiplications. The momentum conservation reduces the
computational cost of the CCD equations to
$n_{\mathrm{o}}n_{\mathrm{u}}^{3}$, where $n_{\mathrm{o}}$
($n_{\mathrm{u}}^{3}$) is the number of occupied (unoccupied) momentum
states. This is a considerable reduction in computational cycles as
compared to the normal cost of the CCD equations which is
$n_{\mathrm{o}}^2n_{\mathrm{u}}^{4}$ \cite{bartlett2007}, and similar
to the reduction of computational cost achieved in the angular
momentum coupled scheme \cite{hagen2008,hagen2010b}.

The coupled-cluster equations~(\ref{eq:t2amp}) are solved numerically
by iteration and yield the matrix elements of $T_2$.  The CCD
energy~(\ref{eq:ccd_energy}) is given in algebraic form by
\[
  E_{\mathrm{CCD}} = E_0 + \frac{1}{4}\sum_{ijab}
  \braket{\mathbf{k}_{i}\mathbf{k}_{j}}{v|\mathbf{k}_{a}\mathbf{k}_{b}}
  \braket{\mathbf{k}_{a}\mathbf{k}_{b}}{t|\mathbf{k}_{i}\mathbf{k}_{j}}.
\]

Below, we will also employ an approximation (denoted as CCD$_{\mathrm
  {ladd}}$) that only uses the particle-particle and hole-hole ladders
in the CCD equations, i.e. 
\begin{eqnarray} \label{eq:t2amp_ladd}
  0 &=&  \braket{\mathbf{k}_{a}\mathbf{k}_{b}}{v|\mathbf{k}_{i}\mathbf{k}_{j}} \nonumber \\
  & + &  P(ab) \sum_c \braket{\mathbf{k}_{b}}{f|\mathbf{k}_{c}} 
  \braket{\mathbf{k}_{a}\mathbf{k}_{c}}{t|\mathbf{k}_{i}\mathbf{k}_{j}} \nonumber \\
  & - &  P(ij)\sum_k  \braket{\mathbf{k}_{k}}{f|\mathbf{k}_{j}} 
  \braket{\mathbf{k}_{a}\mathbf{k}_{b}}{t|\mathbf{k}_{i}\mathbf{k}_{k}} \nonumber \\
  & + & \frac{1}{2}\sum_{cd}\braket{\mathbf{k}_{a}\mathbf{k}_{b}}{v|\mathbf{k}_{c}\mathbf{k}_{d}}
  \braket{\mathbf{k}_{c}\mathbf{k}_{d}}{t|\mathbf{k}_{i}\mathbf{k}_{j}} \nonumber \\
  & + & \frac{1}{2}\sum_{kl}\braket{\mathbf{k}_{a}\mathbf{k}_{b}}{t|\mathbf{k}_{k}\mathbf{k}_{l}}
  \braket{\mathbf{k}_{k}\mathbf{k}_{l}}{v|\mathbf{k}_{i}\mathbf{k}_{j}} \ .
\end{eqnarray}

The CCD$_{\mathrm{ladd}}$ approximation was used in
Ref.~\cite{baardsen2013} within coupled-cluster theory, and a similar
approximation was also employed in other computations of nucleonic
matter, see, e.g., Refs.~\cite{hebeler2010b,hebeler2011}.

Let us also discuss the inclusion of three-body clusters.  When going
beyond the CCD approximation and considering triples excitations, one
might question whether the residual three-body part $w$ can safely be
neglected. After all, three-body forces directly induce excitations of
three-body clusters. Below we will include the residual part $w$ when
considering contributions from triples excitations to the correlation
energy, and study the accuracy of the normal-ordered two-body
approximation in the presence of triples excitations in neutron and
symmetric nuclear matter. Very recently, Binder {\it et al.} employed
chiral interactions softened via the similarity renormalization group
transformation~\cite{bogner2007,roth2012}, studied the effect of
triples corrections in the presence of 3NFs in nuclei such as $^{16}$O
and $^{40}$Ca, and found it to be small~\cite{binder2013}.

The full inclusion of triples in the presence of three-body forces is
demanding and computationally expensive.  Some effects of triples can
be included in the CCD(T) approximation~\cite{raghavachari1985} that
we extend to 3NFs.  In CCD(T) the triples excitation amplitude is
approximated as
\beqa
  \label{eq:ccdt}
  t^{abc}_{ijk}&\equiv& \braket{\mathbf{k}_{a}\mathbf{k}_{b}\mathbf{k}_{c}}{t|\mathbf{k}_{i}\mathbf{k}_{j}\mathbf{k}_{k}}\nonumber\\
&\approx&   \langle \Phi_{ijk}^{abb} \vert (v+w)(1+{\hat T}_2) \vert \Phi_0 \rangle / \epsilon^{ijk}_{abc}.  
\eeqa
Here
\beq
\epsilon^{ijk}_{abc} \equiv f^i_i + f^j_j +f_k^k -f_a^a-f^b_b-f^c_c \ .
\eeq
The CCD(T) correction to the energy is 
\begin{equation}
  \Delta E_{\mathrm{CCD(T)}}  =  \frac{1}{36}\sum_{ijkabc} \left|t^{abc}_{ijk}\right|^2 / \epsilon^{ijk}_{abc}. 
  \label{eq:ccdt_energy}
\end{equation}

Employing the triples amplitude~(\ref{eq:ccdt}) with the inclusion of
$w$ yields the energy correction $\Delta E_{\mathrm{CCD(T)}}$.  We
also consider the following approximations.  Neglecting the residual
three-body part $w$ yields the normal-ordered two-body approximation
to the CCD(T) energy correction, denoted as $\Delta
E_{\mathrm{CCD(T:~w=0)}}$. Omitting the term $wT_2$ in
Eq.~(\ref{eq:ccdt}) gives the energy correction $\Delta
E_{\mathrm{CCD(T:}~wT_2=0)}$. Note that the numerically expensive term
$wT_2$ in Eq.~(\ref{eq:ccdt}) consist of three distinct diagrams in
which one sums over $pp$, $hh$ and $ph$ intermediate states,
respectively.  Below we will investigate the contributions of these
three diagrams to the CCD(T) energy correction in neutron and
symmetric nuclear matter.

\subsection{Ladder approximation in a partial-wave basis}
 In Ref.~\cite{baardsen2013}, the ladder approximation of the
 coupled-cluster equations for nuclear matter is presented in an
 alternative formulation. Historically, the equations for nuclear
 matter, for example in the hole-line approximation~\cite{day1978},
 have often been expressed explicitly in a partial-wave basis
 \cite{day1981b,haftel1970,suzuki2000}. Similarly, in the method
 presented in Ref.~\cite{baardsen2013}, the ladder approximation is
 formulated in a partial-wave basis, assuming that the thermodynamic
 limit is reached and therefore using integrals over relative and
 center-of-mass momenta. In the partial-wave expanded equations, the
 Pauli exclusion operators are treated exactly, using a technique
 introduced for the Brueckner-Hartree-Fock approximation by Suzuki
 \emph{et al.} \cite{suzuki2000}. Apart from the truncation in partial
 waves, the only approximation in this method is in the
 single-particle potentials, where an angular-average approximation
 was used for the laboratory momentum
 argument~\cite{brueckner1958,baardsen2013}.

\subsection{Finite size effects}
\label{finite}

We would like to quantify the error due to finite size effects and the
accuracy of our coupled-cluster calculations of neutron and nuclear
matter.  Using periodic boundary conditions (PBC) one could increase
the number of particles in the box until convergence to the
thermodynamic limit is reached. However, due to variations of the
shell effects at different closed shell configurations, there is no
guarantee that increasing the number of particles will lead to a
systematic and smooth convergence to the thermodynamic limit.
Furthermore, the computational cost of many-body methods such as the
AFDMC and coupled-cluster methods increases rapidly with increasing
particle number, and one would therefore like to employ a method that
controls finite size effects already for modest particle numbers. This
can be achieved with averaging over phases of Bloch waves that
correspond to different boundary
conditions~\cite{gros1992,gros1996,lin2001}.

Consider a free particle in a box of size $L$ subject to twisted
boundary condition, that is,  the wave function with momentum $k$ fulfills
the condition for so-called Bloch waves, namely,
$\psi_k(x+L)=e^{i\theta}\psi_k(x)$. By
averaging over the twist angle $\theta$, shell effects can be
eliminated for free Fermi systems~\cite{gros1992}, and they are much
suppressed for interacting systems~\cite{gros1996,lin2001}. In
this way, one obtains a much more systematic and smooth convergence
towards the thermodynamic limit.  The twisted boundary conditions are
defined by
\[
k_{n_i} = \frac{(2\pi n_i+\theta_i)}{L}, \:\: n_i = 0, \pm 1, \ldots
\pm n_{\mathrm{max}}, \:\: i = x,y,z,
\]
with the twist angle $\theta \in [0,\pi]$ for systems with
time-reversal invariance~\cite{lin2001}. This amounts to letting the
particles pick up a complex phase when they wrap
around the boundary of the cubic box. By integrating or averaging over
a finite number of twists in each $x,y,z$ direction we obtain the
twist-averaged boundary conditions (TABC). In our implementation of
TABC we integrate over the twist angles $\theta$ using a finite number
of Gauss-Legendre quadrature points in $[0,\pi]$. Note that $\theta=0$
($\theta = \pi$) corresponds to (anti-)periodic boundary conditions.

\begin{figure}[hbt]
\includegraphics[width=0.9\columnwidth]{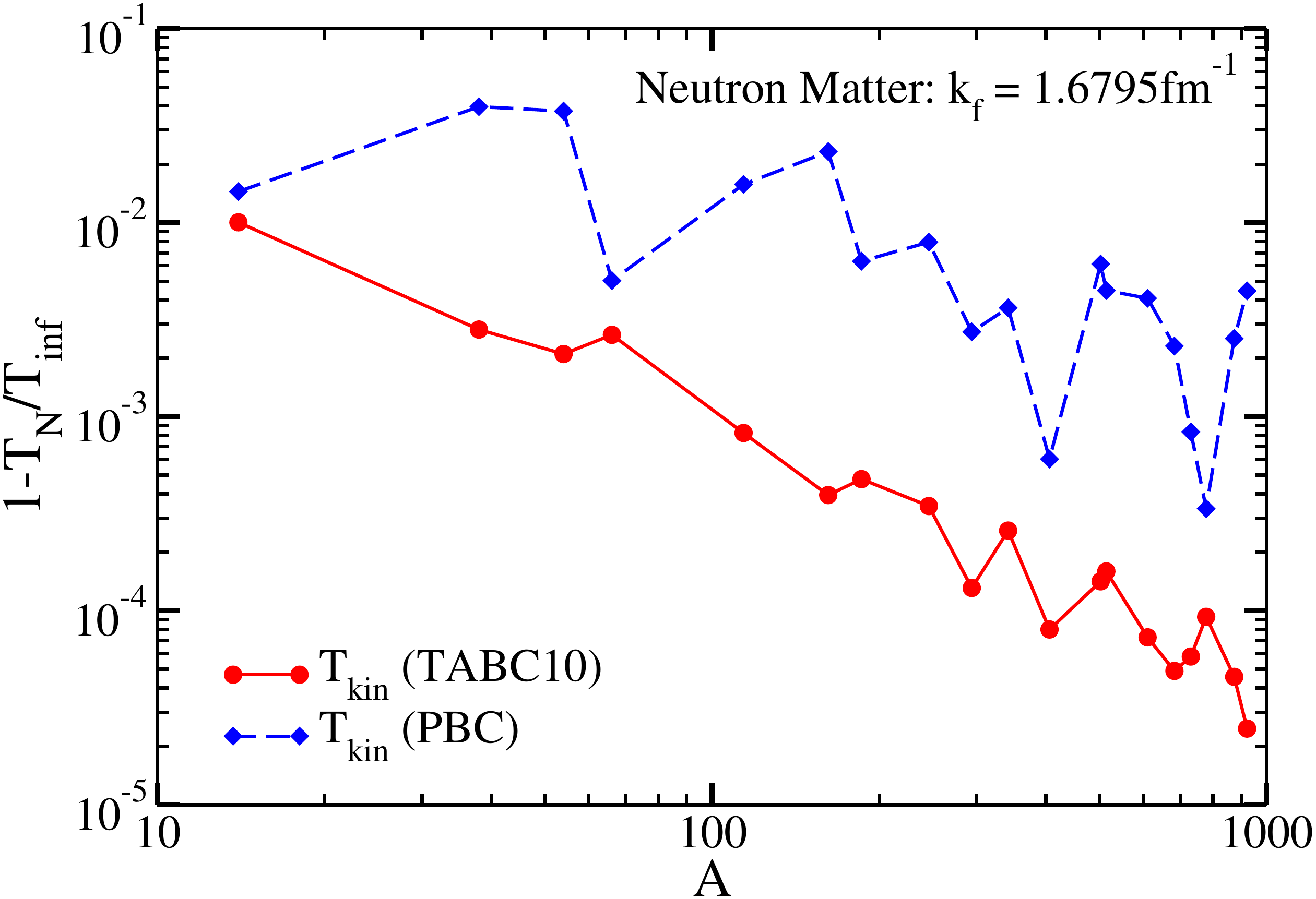}
\caption{(Color online) Relative finite-size corrections for the
  kinetic energy in pure neutron matter at the Fermi momentum
  $k_F=1.6795\mathrm{fm}^{-1}$ vs. the neutron number $A$. TABC10 are
  twist-averaged boundary conditions with 10 Gauss-Legendre points in
  each spatial direction.}
\label{fig:kinetic_finitesize}
\end{figure}

In order to quantify the finite size effects using PBC and TABC we
compute the kinetic and potential energy contribution to the
Hartree-Fock energy for several closed shell configurations ranging
from tens to several hundreds of nucleons, and compare with the
thermodynamic limit for these quantities. In
Fig.~\ref{fig:kinetic_finitesize} we show the relative error of the
kinetic energy in pure neutron matter for the Fermi momentum
$k_F=1.6795~\mathrm{fm}^{-1}$ computed using standard PBC and TABC. We
used 10 Gauss-Legendre points for the twist angle $\theta_i$ of the $i
= x,y,z$ direction in the integration interval $[0,\pi]$. Clearly, we
obtain a much faster and smoother convergence to the thermodynamic
limit using TABC. Generally we get about an order of magnitude
reduction in the relative error when using TABC as compared to PBC.
Finite size effects are particularly small for PBC and $N=66$
neutrons. This was also seen in AFDMC
calculations~\cite{gandolfi2009}.

\begin{figure}[hbt]
\includegraphics[width=0.9\columnwidth]{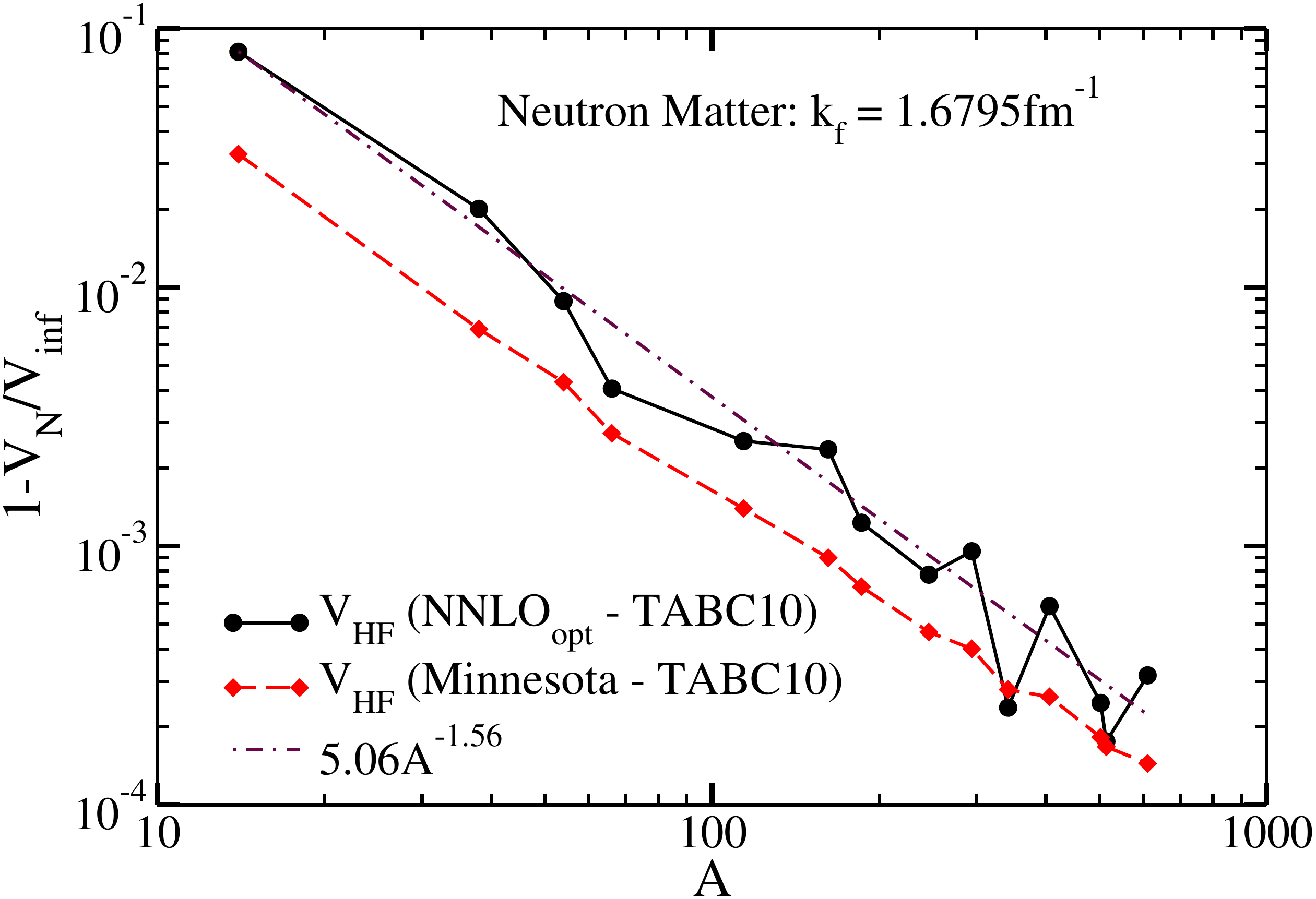}
\caption{(Color online) Relative finite-size corrections for the
  Hartree-Fock energy of the NNLO$_\mathrm{opt}$ (full line) and
  Minnesota (dashed line) potentials in pure neutron matter at the
  Fermi momentum $k_F=1.6795\mathrm{fm}^{-1}$ vs. the neutron number
  $A$. TABC10 are twist-averaged boundary conditions with 10
  Gauss-Legendre points in each spatial direction. The dashed-dotter
  line shows a power law fit to the NNLO$_\mathrm{opt}$ results. }
\label{fig:pnm_hf_finitesize}
\end{figure}

Figure~\ref{fig:pnm_hf_finitesize} shows the relative error of the
potential energy to the Hartree-Fock energy in pure neutron matter for
the Fermi momentum $k_F=1.6795~\mathrm{fm}^{-1}$ computed with TABC.
We compute the potential energy from NNLO$_\mathrm{opt}$ and from the
Minnesota potential. We see that the finite size effects in the
potential energy are comparable to the finite size effects in the
kinetic energy shown in Fig.~\ref{fig:kinetic_finitesize}. We note
that finite size effects vanish as the power law $A^{-1.56}$ in the
neutron number $A$.

\begin{figure}[hbt]
\includegraphics[width=0.9\columnwidth]{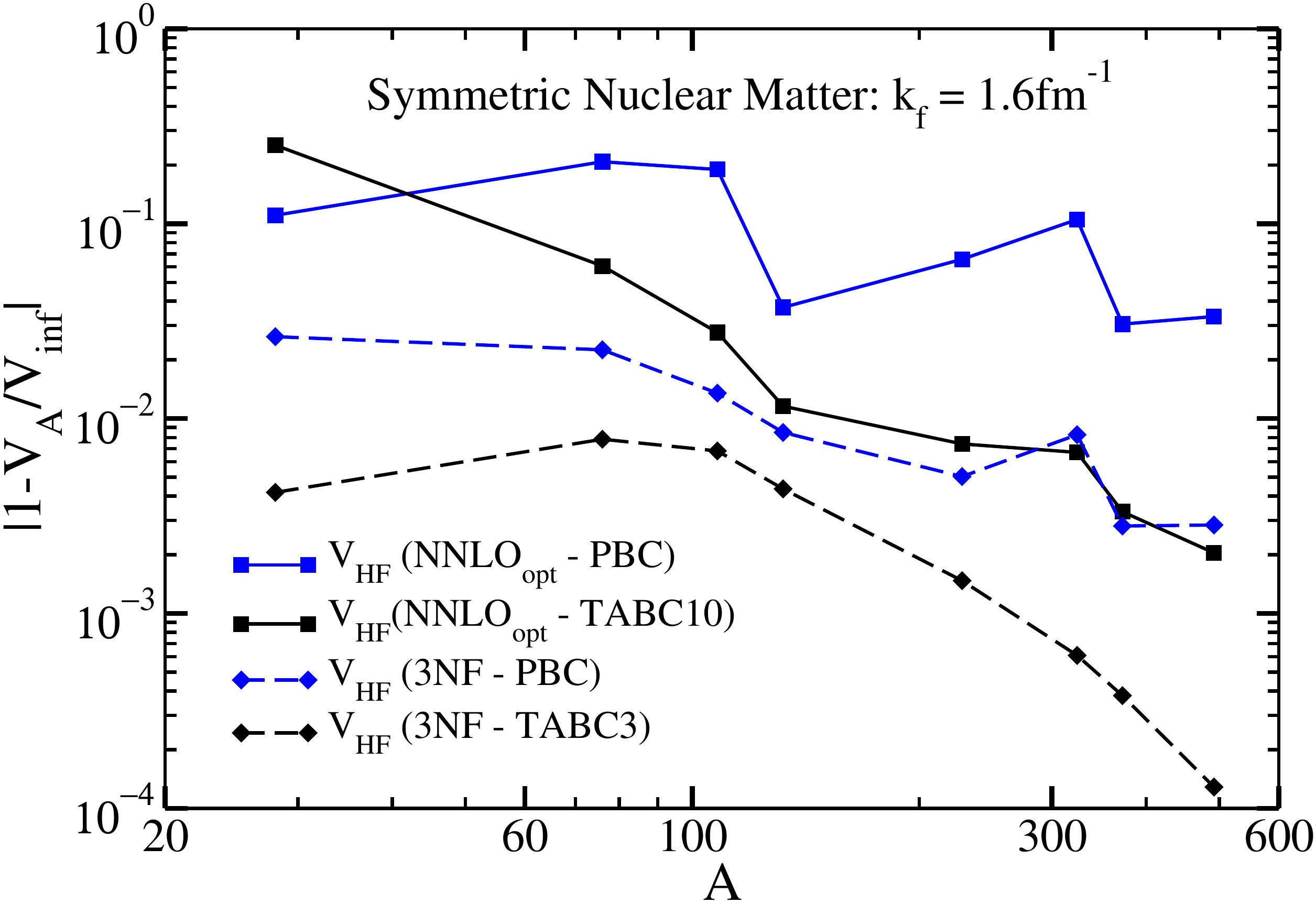}
\caption{(Color online) Relative finite-size corrections for the
  Hartree-Fock energy of the $NN$ potential NNLO$_\mathrm{opt}$ and
  the 3NF potential in symmetric nuclear matter at the Fermi momentum
  $k_F=1.6\mathrm{fm}^{-1}$ vs. the nucleon number $A$. PBC: periodic
  boundary conditions. TABC10 and TABC3 are twist-averaged boundary
  conditions with 10 and 3 Gauss-Legendre points in each spatial
  direction, respectively.}
\label{fig:snm_hf_finitesize}
\end{figure}

Finally, we would also like to assess the finite-size effects in
symmetric nuclear matter. In Fig.~\ref{fig:snm_hf_finitesize} we show
the relative error of the potential energy to the Hartree-Fock energy
in symmetric nuclear matter for the Fermi momentum
$k_F=1.6~\mathrm{fm}^{-1}$ computed using PBC and TABC. We consider
the Hartree-Fock potential energy contribution from the
nucleon-nucleon interaction NNLO$_\mathrm{opt}$ and the 3NF at order
NNLO separately.  In particular it is seen that the relative error in
the potential energy contribution from the 3NF is about an order of
magnitude smaller than the relative error coming from the
nucleon-nucleon interaction alone using both PBC and TABC. In the case
of symmetric nuclear matter there is no systematic convergence trend
using PBC, and for 132 nucleons the relative error for PBC is around
$\sim 4\%$, while using TABC the error is reduced to $\sim 1\%$. It is
interesting to note that finite size effects for NNLO$_\mathrm{opt}$
with TABC decrease as $A^{-1.59}$ with increasing nucleon number $A$.
This exponent is similar to the exponent found in neutron matter (see
Fig.~\ref{fig:pnm_hf_finitesize}).

Coupled-cluster calculations of nucleonic matter using TABC are very
expensive.  Using 10 twist angles in each direction requires $10^3$
coupled-cluster calculations, although symmetry considerations can
reduce this number considerably. In Ref.~\cite{lin2001} it was shown
that one can find a specific choice of twist angles (known as special
points), in which the Hartree-Fock energy exactly corresponds to the
Hartree-Fock energy in the thermodynamic limit. In the following we
compute these special points for neutron and nuclear matter using both
$NN$ interactions and 3NFs, and compare with calculations using PBC
and TABC.

\subsection{Benchmarks}
\begin{figure}[htb]
\includegraphics[width=0.9\columnwidth]{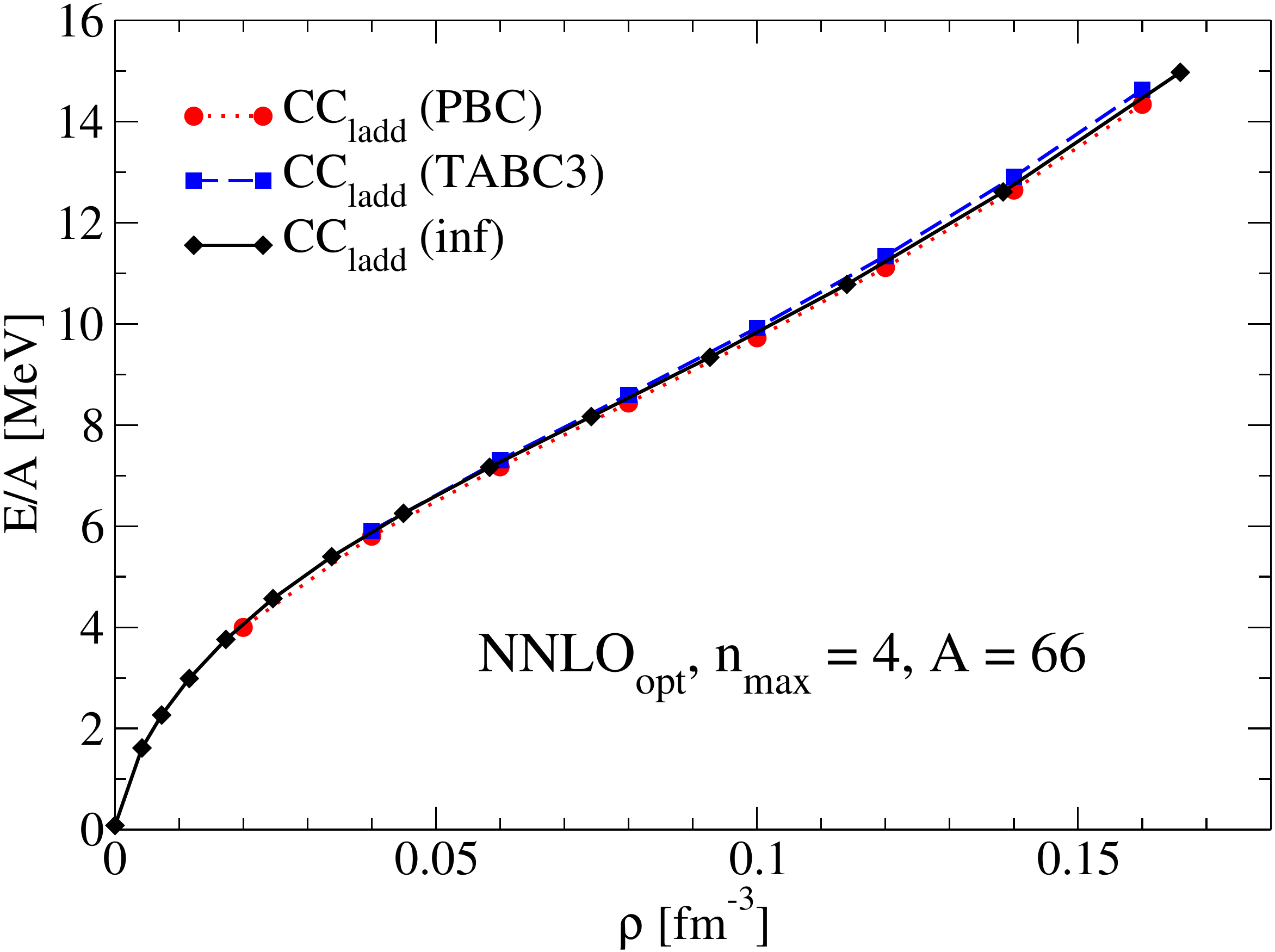}
\caption{(Color online) Energy per particle of neutron matter for
  NNLO$_{\rm opt}$ computed in the CCD$_{\rm lad}$ approximation with
  periodic boundary conditions (circles), twist-averaged boundary
  conditions (squares), and for infinite matter (diamonds). The latter
  results are from Ref.~\cite{ekstrom2013}.  The calculations used
  $A=66$ neutrons and $n_{\mathrm max} =4$. }
\label{fig_bc_neutron}
\end{figure}

It is interesting to compare the results for various boundary
conditions with the infinite matter results by Baardsen {\it et
  al.}~\cite{baardsen2013}.  Figure~\ref{fig_bc_neutron} shows the
CCD$_{\rm ladd}$ results for neutron matter computed with the
nucleon-nucleon potential NNLO$_{\rm opt}$. In a finite system, the
neutron number $N=66$ is very close to the infinite matter results for
both periodic and twist-averaged boundary conditions.

\begin{figure}[htb]
\includegraphics[width=0.9\columnwidth]{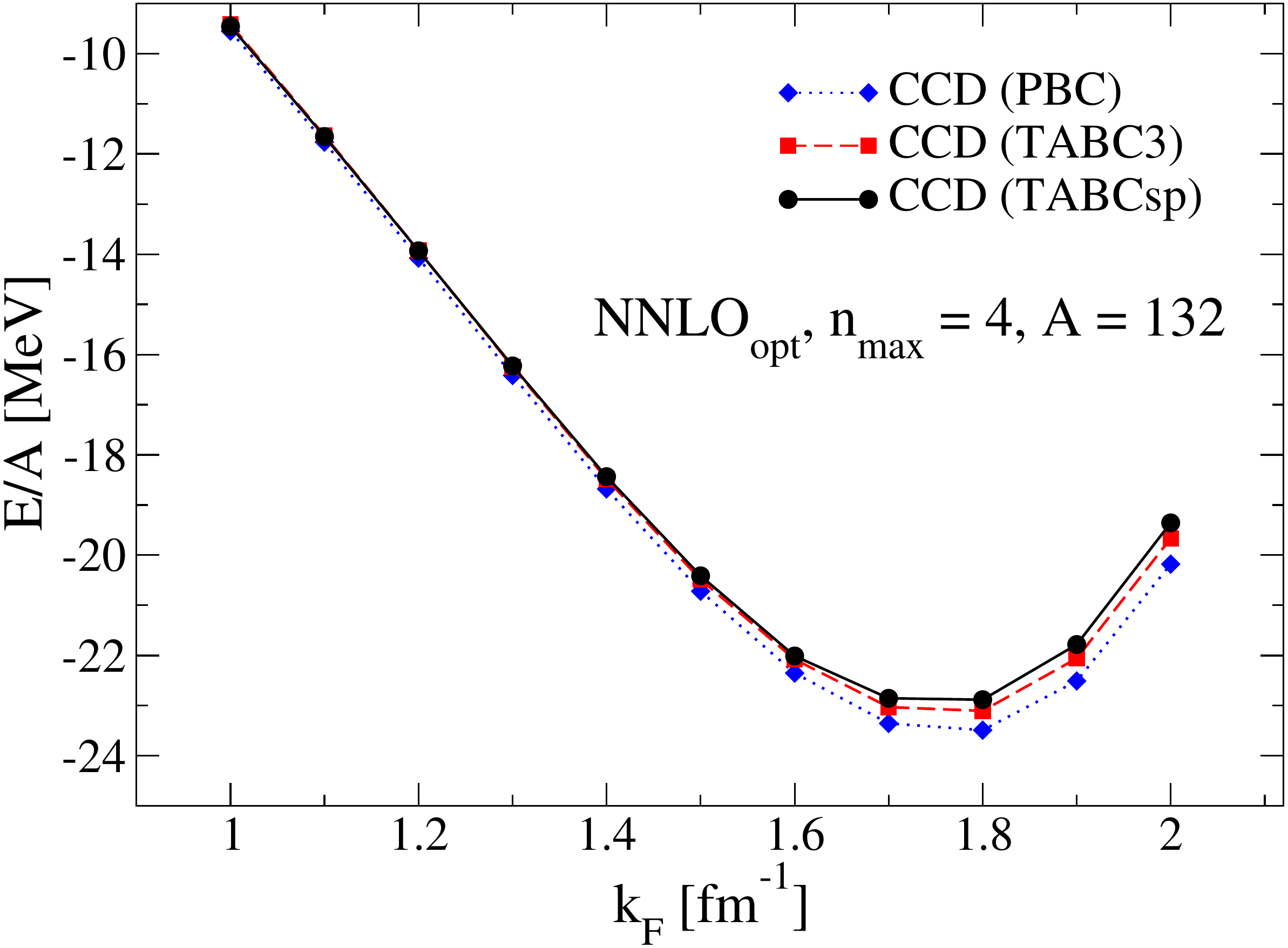}
\caption{(Color online) Energy per particle of symmetric nuclear matter for
  NNLO$_{\rm opt}$ computed in the CCD approximation with periodic
  boundary conditions (diamonds), twist-averaged boundary conditions
  (squares), and with a special point and twisted boundary conditions
  (circles). The calculations used $A=132$ nucleons and $n_{\mathrm max}
  =4$. }
\label{fig_bc_nuclear}
\end{figure}

For symmetric nuclear matter, the CCD results are more sensitive to
the choice of the boundary conditions, with results shown in
Fig.~\ref{fig_bc_nuclear}.  At higher Fermi momenta ($k_F >
1.6~\mathrm{fm}^{-1}$), the energy per nucleon for periodic boundary
conditions differs by $\sim 0.5$ MeV from the result obtained with
twist-averaged boundary conditions. A calculation with a special point
in the twist is very close to the twist-averaged results. However, for
Fermi momenta $k_F < 1.6~\mathrm{fm}^{-1}$, the difference between the
PBC and TABC is less than $200$~keV per nucleon.

\begin{figure}[htb]
\includegraphics[width=0.9\columnwidth]{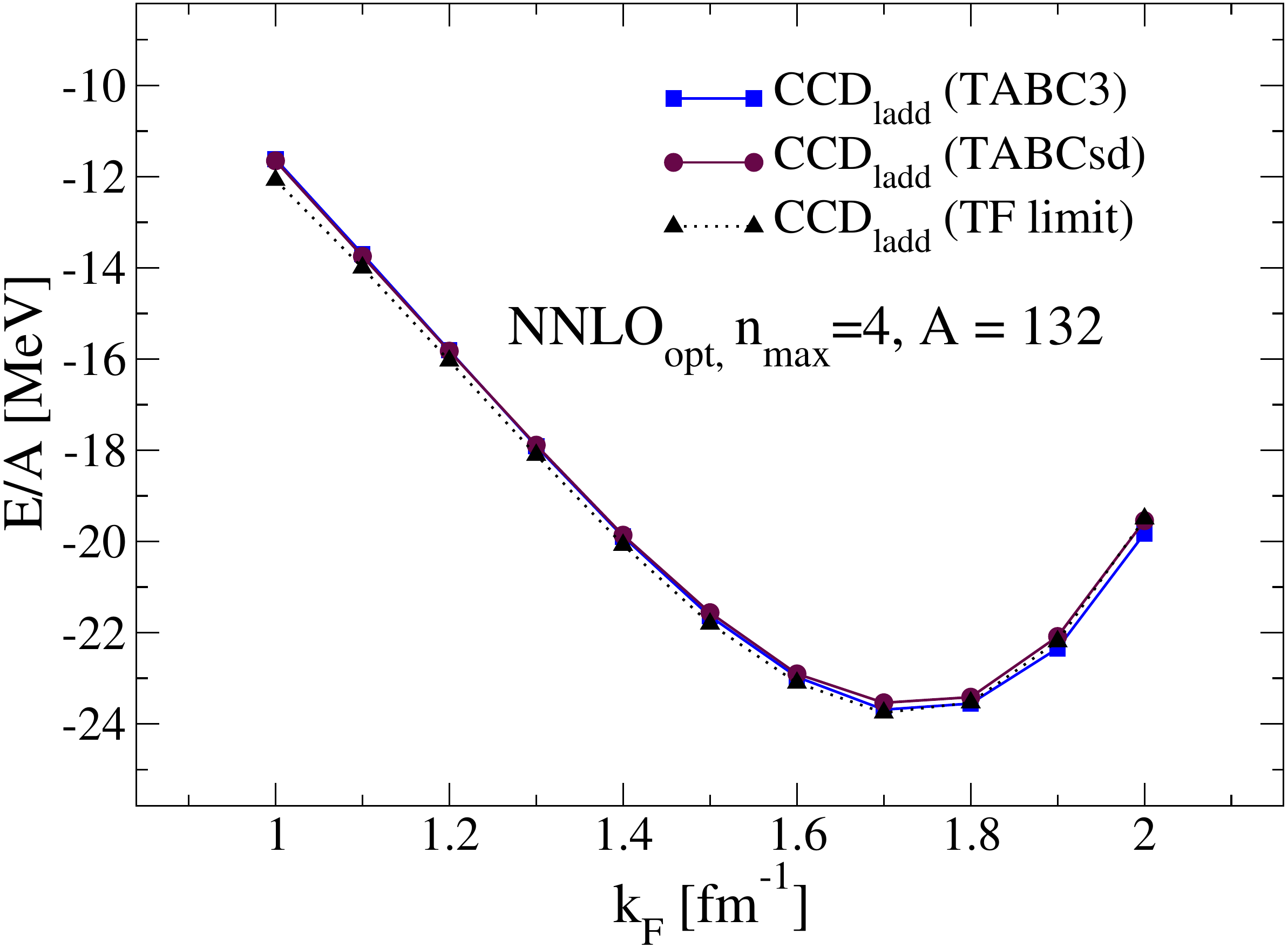}
\caption{(Color online) Energy per particle of symmetric nuclear
  matter for NNLO$_{\rm opt}$ computed in the CCD$_{\mathrm ladd}$
  approximation in the thermodynamic limit using partial-wave
  expansion (triangles) (partly adapted from
  Ref.~\cite{baardsen2013}), with twist averaged boundary conditions
  (squares), and with a special point and twisted boundary conditions
  (circles). The calculations used $A=132$ nucleons and $n_{\mathrm
    max} =4$. }
\label{fig_bc_nuclear2}
\end{figure}

Figure~\ref{fig_bc_nuclear2} compares nuclear matter results
calculated in the ladder approximation with the CC$_{\rm ladd}$ of
Ref.~\cite{baardsen2013}. The latter were obtained by taking the
thermodynamic limit in the relative and center-of mass frame and by
summing over partial waves.  The summation over intermediate
particle-particle and hole-hole configurations is performed with an
exact Pauli operator, while the single-particle energies are computed
using an angle-averaging procedure, see Ref.~\cite{baardsen2013} for
further details.  For these results, the angle-average approximation,
together with a truncation in the number of partial waves included,
represent the sources of possible errors in the thermodynamic limit.
It is therefore very satisfactory that the results from different
methods are close to each other.

Let us also consider a simple potential model and benchmark the
results of our coupled-cluster calculations against virtually exact
results from the auxiliary field diffusion Monte Carlo (AFDMC)
method~\cite{schmidt1999}.  The Minnesota
potential~\cite{thompson1977} is a semi-realistic nucleon-nucleon
interaction that can be solved accurately with AFDMC.  It depends only
on the relative momenta and spin, but lacks spin-orbit or tensor
contributions. The matrix elements of this potential are real numbers.
For the benchmark we employ periodic boundary conditions, $A=66$
neutrons, and $n_{\mathrm{max}}=6$.

\begin{figure}[htb]
\includegraphics[width=0.9\columnwidth]{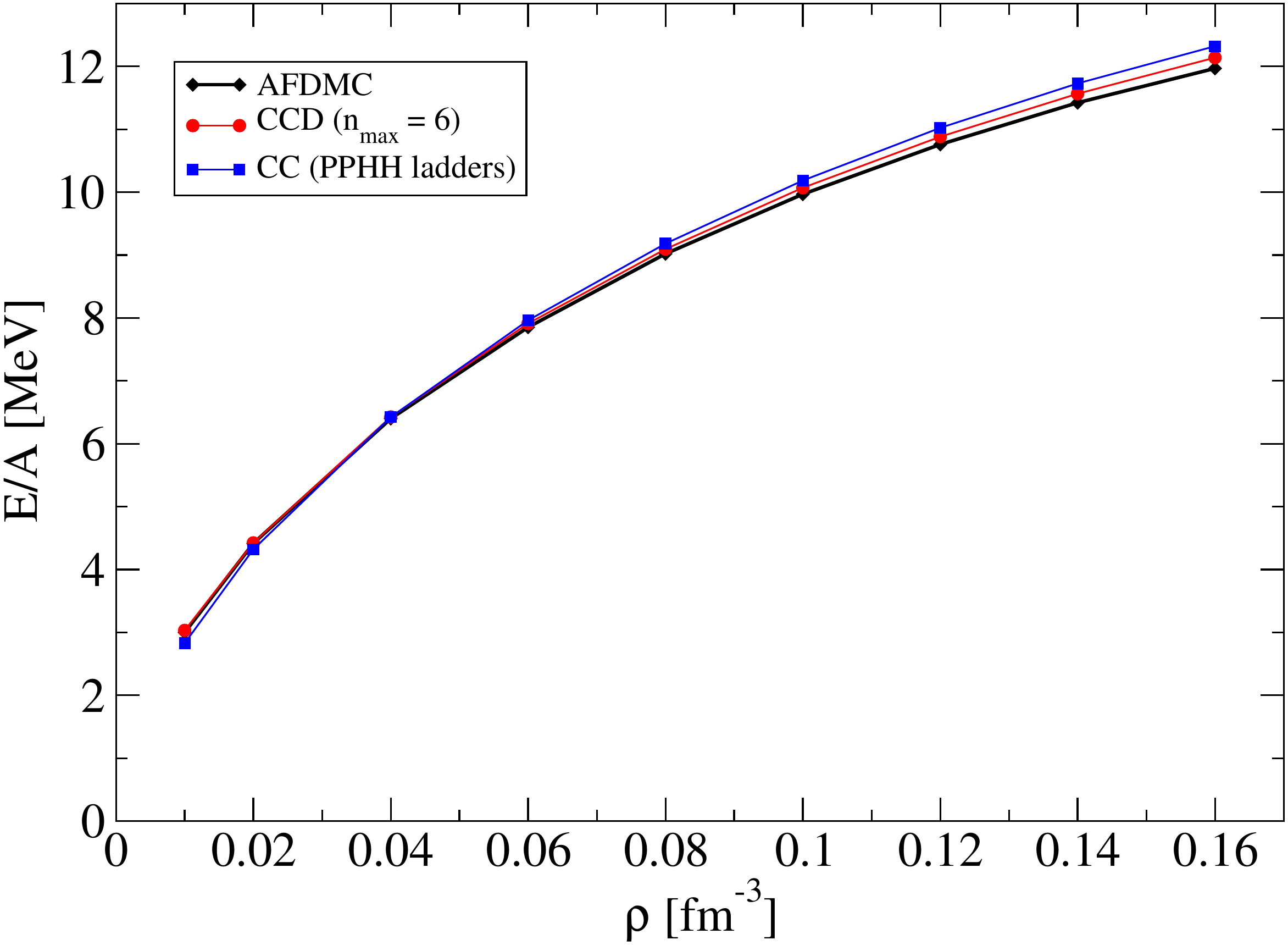}
\caption{(Color online) Energy per particle of neutron matter, computed
  with the Minnesota potential. Diamonds: AFDMC, circles: CCD,
  squares: CCD limited to $pp$ and $hh$ ladders.}
\label{fig_afdmc}
\end{figure}

Figure~\ref{fig_afdmc} compares the energy per neutron of our lattice
CCD results (circles), and our CCD$_{\mathrm{ladd}}$ in the
thermodynamic limit, see Ref.~\cite{baardsen2013}, to
the AFDMC benchmark. Overall, the agreement is good between all
methods. As expected, the CCD results are more accurate than the
CCD$_{\mathrm{ladd}}$ approximation.

\begin{figure}[htb]
\includegraphics[width=0.9\columnwidth]{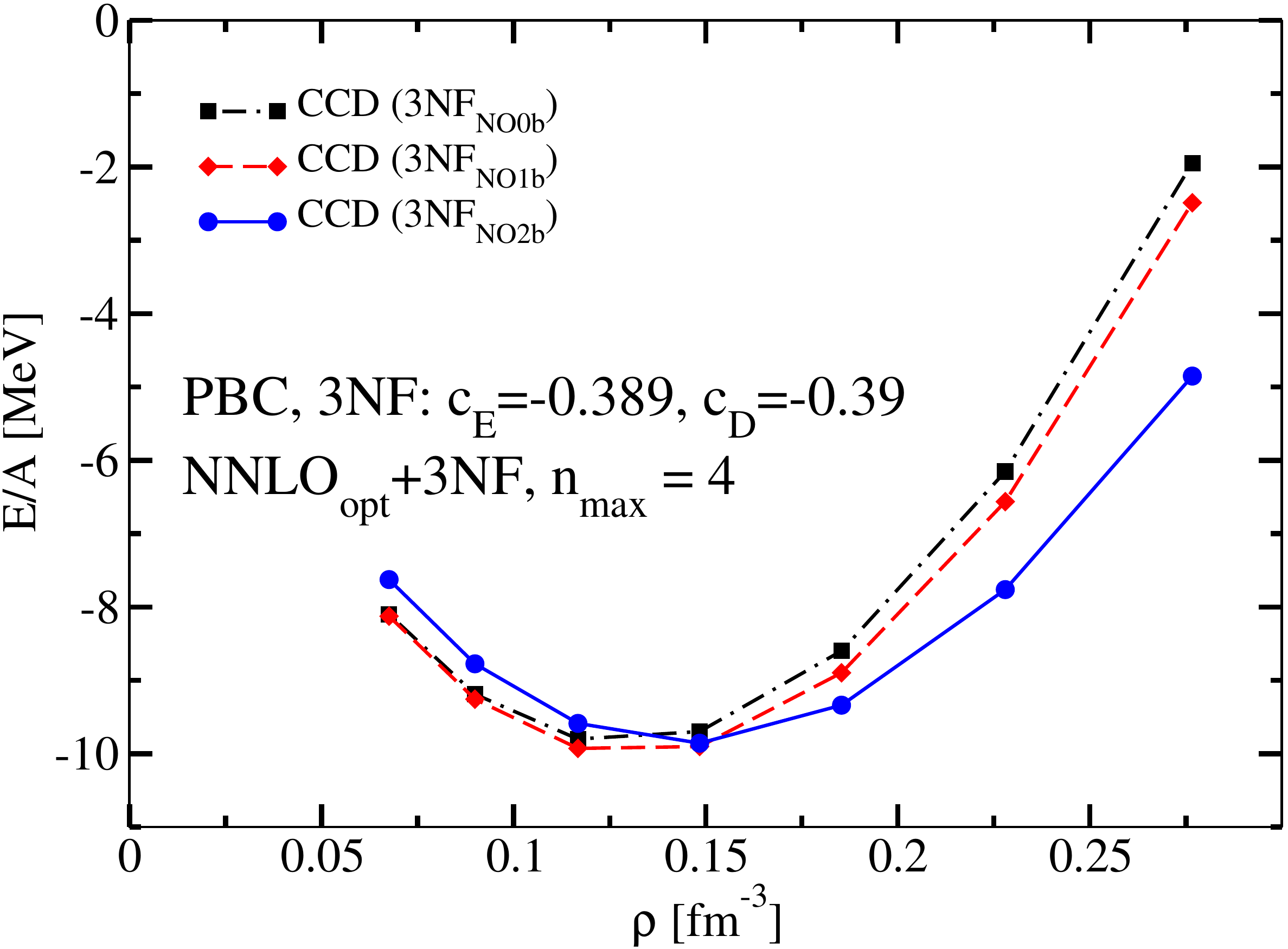}
\caption{(Color online) Energy per particle of symmetric nuclear
  matter computed in the CCD approximation with NNLO$_{opt}$ + 3NF
  (c$_{\mathrm E}=$-0.389, c$_{\mathrm D}=$-0.39). The 3NF is included
  in the 0-body (black dashed-dotted line), 1-body (red dashed line),
  and in the two-body (blue solid line) normal-ordered
  approximations. The calculations used $A=132$ nucleons and
  $n_{\mathrm max} =4$. }
\label{fig_3NFno}
\end{figure}

Finally, we turn to 3NFs. The inclusion of 3NFs -- even in the
normal-ordered approximation -- is still numerically expensive due to
the large number of required matrix elements.  We also study different
approximations for 3NFs, and compare the results for symmetric nuclear
matter when 3NFs only enter in the normal-ordered approximation as
0-body, 1-body, or up to 2-body forces. Figure~\ref{fig_3NFno} clearly
shows that normal-ordered 2-body forces are relevant.

\section{Results for chiral interactions}\label{sec:results} 

In this Section, we present our results for coupled-cluster
computations of neutron matter and symmetric nuclear matter. As shown
in the previous Section, the finite size effects (and the differences
between PBC and TABC) are small for $A=66$ neutrons and $A=132$
nucleons when calculating neutron matter and symmetric nuclear matter,
respectively. For this reason, many of the expensive calculations
involving 3NFs are only performed with PBC at these specific particle
numbers.

\subsection{Neutron matter} 
\begin{figure}[hbt]
\includegraphics[width=0.9\columnwidth]{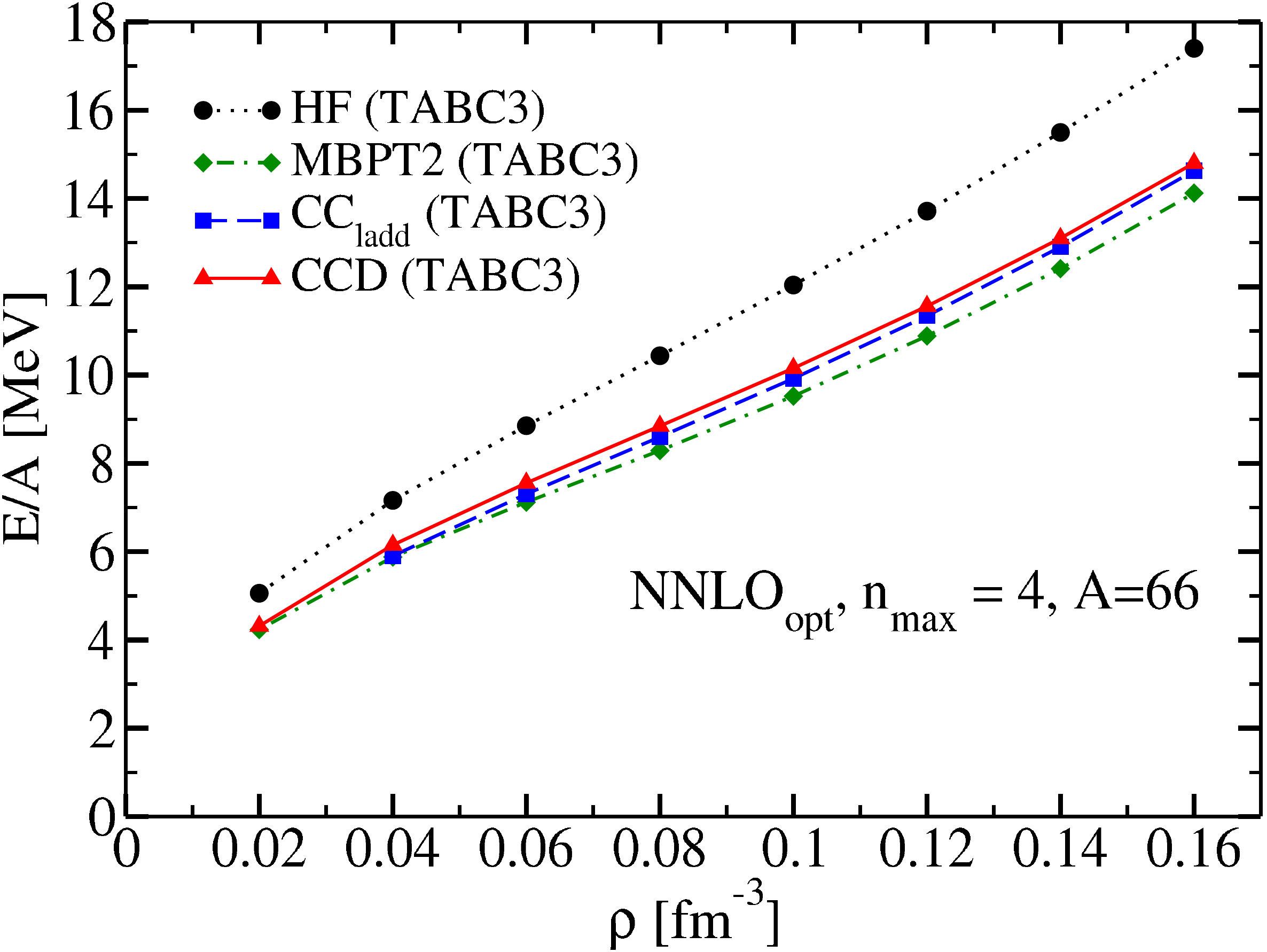}
\caption{(Color online) Energy per particle in neutron matter with
  NNLO$_\mathrm{opt}$ (NN only). The black dashed line is Hartree-Fock
  (HF), the green dashed-dotted line is second-order many-body
  perturbation theory (MBPT2), the blue dashed line is coupled-cluster
  doubles ladder approximation (CCD$_{\mathrm{ladd}}$), and the red
  solid line is coupled-cluster doubles (CCD).  The
  calculations used $A=66$ neutrons, $n_{\mathrm max}=4$, and TABC3.}
\label{fig:pnm_nnlo}
\end{figure}

Figure~\ref{fig:pnm_nnlo} shows the energy per neutron as a function
of density based on $NN$ interactions alone and compares various
many-body methods. The employed $NN$ interaction NNLO$_{\rm opt}$ is
perturbative in neutron matter, with second-order many-body
perturbation theory (MBPT2), CCD and CCD$_{\rm ladd}$ giving similar
results that differ by less than 1~MeV per neutron at nuclear
saturation density.

\begin{figure}[hbt]
\includegraphics[width=0.9\columnwidth]{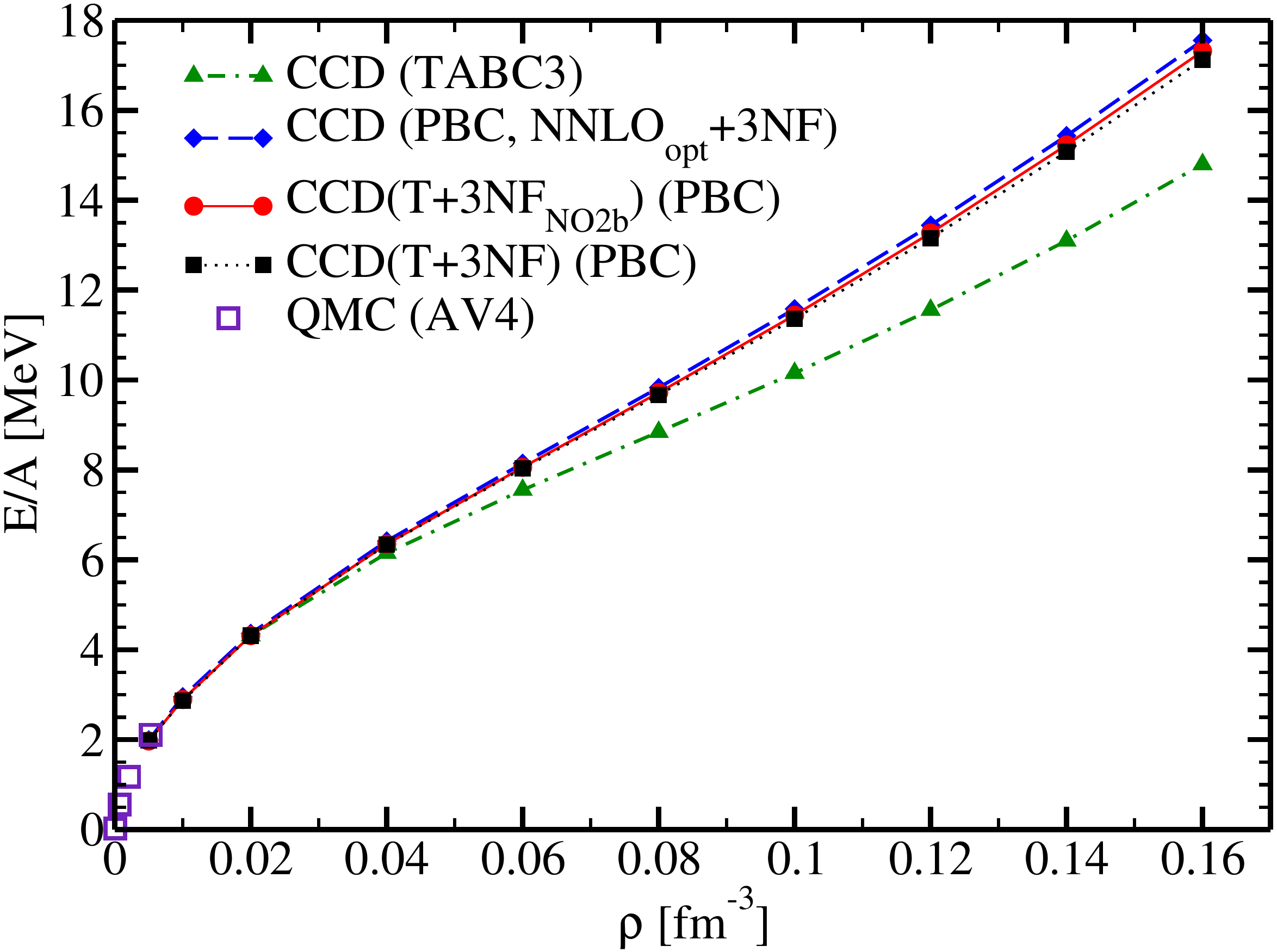}
\caption{(Color online) Energy per particle in neutron matter with
  NNLO$_\mathrm{opt}$ (NN only) and with inclusion of 3NF computed in
  the CCD and CCD(T) approximations. The 3NF LECs are given by $c_E =
  -0.389$ and $c_D = -0.39$. The calculations used $A=66$ neutrons,
  $n_{\mathrm max}=4$, and PBC and TABC.}
\label{fig:pnm_nnlo3nf}
\end{figure}

Figure~\ref{fig:pnm_nnlo3nf} shows the effect of 3NFs in CCD
calculations of the EoS for neutron matter. We consider several
approximations involving 3NFs, and it is seen that they yield very
similar results.  We note that three-nucleon forces act repulsively.
The results for neutron matter reported here are consistent with the
recent calculations of Krueger \emph{et al.}~\cite{krueger2013}, and
our results for the EoS fall within their NNLO uncertainty band.  The
CCD calculation that includes the normal-ordered 3NFs is shown as
diamonds. Triples corrections that are limited to the inclusion of up
to two-body terms from the normal-ordered 3NF are shown as circles,
while triples corrections that include also the residual 3NF are shown
as squares.  For neutron matter, the effects of triples are small and
account for about 0.3~MeV per neutron at high densities, and the
residual 3NFs contribute little to the triples corrections.

\subsection{Nuclear matter} 
\begin{figure}[htb]
\includegraphics[width=0.9\columnwidth]{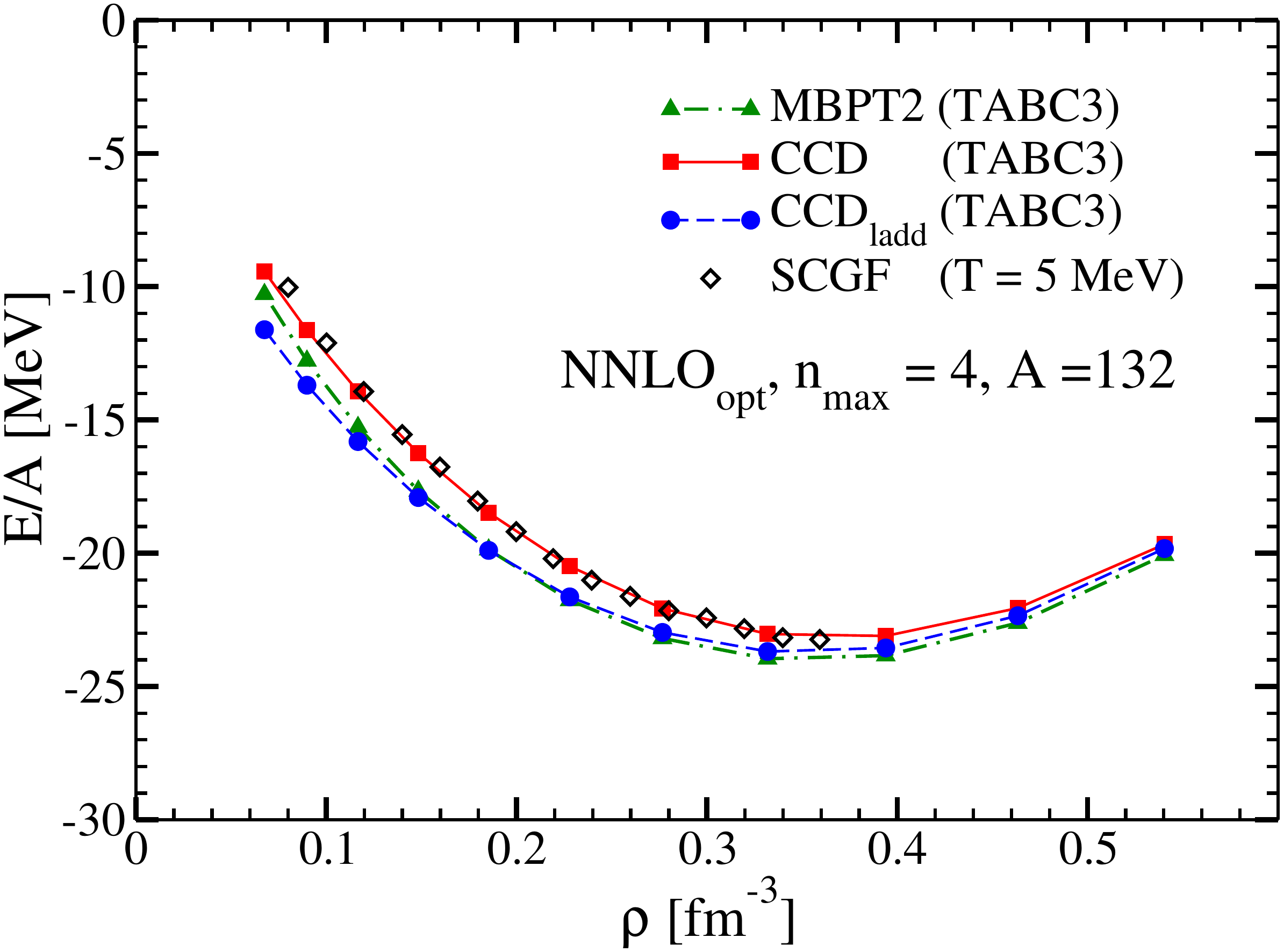}
\caption{(Color online) Energy per particle in symmetric nuclear
  matter with NNLO$_\mathrm{opt}$ (NN only) computed in the MBPT2
  (triangles with green dashed-dotted line), CCD$_{\mathrm{ladd}}$
  (circles with blue dashed line), and CCD (squares with solid red
  line) approximations. The calculations used 132 nucleons,
  $n_{\mathrm max}=4$, and PBC. Diamonds are results from
  self-consistent Green's function (SCGF) at the finite temperature
  $T=5$~MeV, taken from Ref.~\cite{carbone2013}.}
\label{fig:snm_nnlo}
\end{figure}

In this Subsection we perform coupled-cluster calculations of
symmetric nuclear matter using chiral $NN$ and 3NF interactions at
NNLO.  Figure~\ref{fig:snm_nnlo} shows the energy per nucleon in
symmetric nuclear matter for a wide range of densities computed in
MBPT2, the CCD$_{\mathrm{ladd}}$, and in the CCD approximation with
the $NN$ potential NNLO$_{\mathrm{opt}}$. In these calculations we
used $A=132$ nucleons, $n_{\mathrm{max}}=4$, and TABC based on $3^3$
angles. We observe that the saturation point is at a too large
density, and we get a considerable overbinding. These results for
NNLO$_{\mathrm{opt}}$ are in good agreement with the recent
self-consistent Green's function (SCGF) calculations of nuclear matter
\cite{carbone2013}, and the CCD$_{\mathrm{ladd}}$ calculations of
Ref.~\cite{baardsen2013}. The difference between MBPT2 and CCD is
considerable, indicating that nuclear matter for the NNLO$_{\mathrm
  {opt}}$ chiral interaction is not perturbative.  The difference
between the CCD$_{\mathrm{ladd}}$ approximation and the full CCD
calculations is around 1~MeV per nucleon around saturation density. We
can conclude that -- in contrast to neutron matter -- for nuclear
matter and the NNLO$_{\mathrm{opt}}$ interaction (which is rather
soft), non-linear terms in the $T_2$ amplitude and particle-hole
excitations yield non-negligible contributions. We note also that the
coupled-cluster calculations are difficult to converge for Fermi
momenta smaller than about $0.8~\mathrm{fm}^{-1}$. This is presumably
due to the clustering of nuclear matter at low
densities~\cite{horowitz2006}.

\begin{figure}[htb]
\includegraphics[width=0.9\columnwidth]{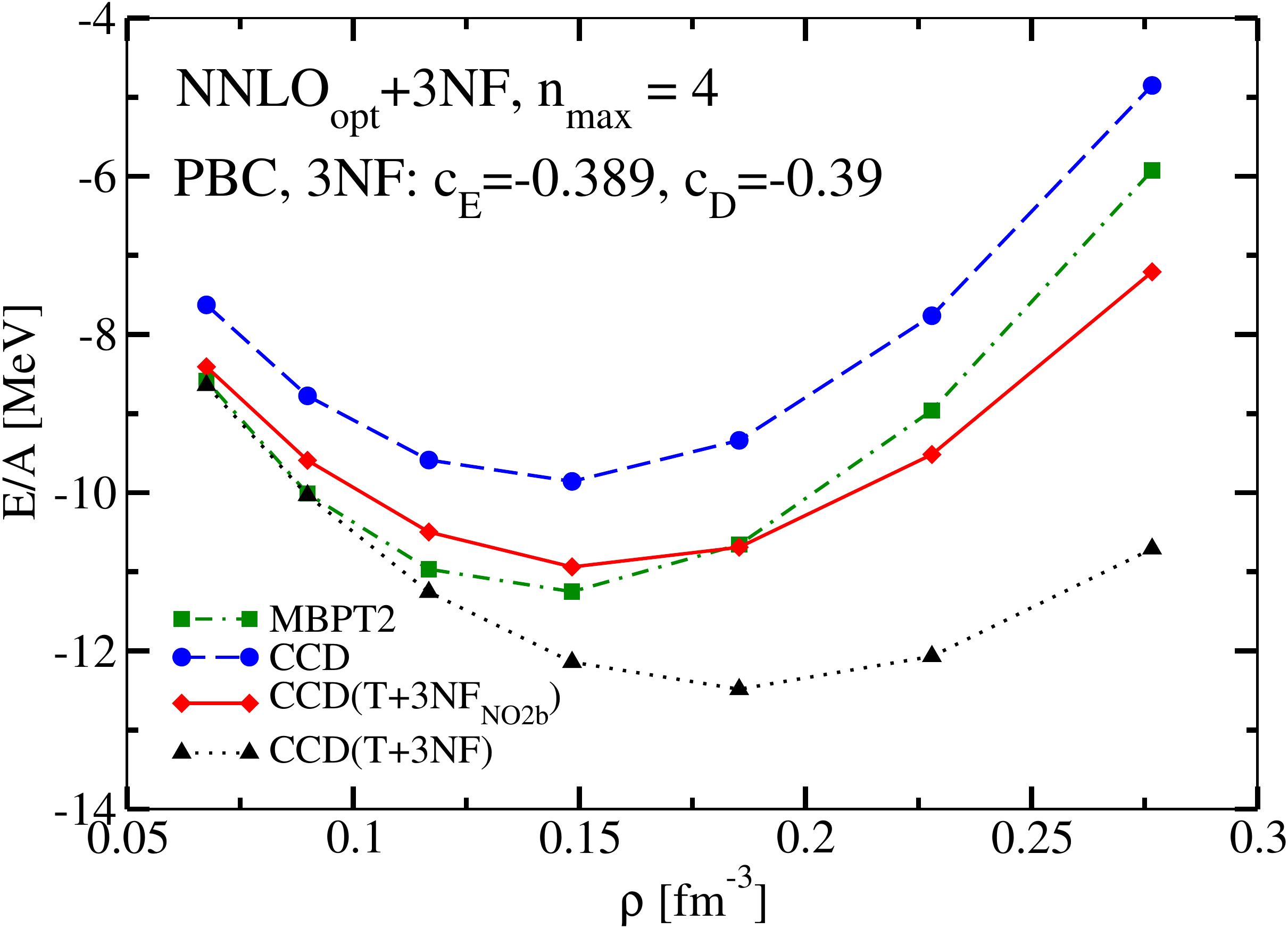}
\caption{(Color online) Energy per particle in symmetric nuclear
  matter with NNLO$_\mathrm{opt}$ and 3NF computed in the MBPT2
  (squares with green dashed-dotted line), CCD (circles with blue
  dashed line), and CCD(T) with 3NF in the normal-ordered two-body
  approximation (diamonds with solid red line), and including the
  residual 3NF in leading order (triangles with dotted black
  line). The 3NF LECs are given by $c_E = -0.389$ and $c_D =
  -0.39$. The calculations used 132 nucleons, $n_{\mathrm max}=4$, and
  PBC.}
\label{fig:snm_nnlo3nf}
\end{figure}

Let us turn to 3NFs. Figure~\ref{fig:snm_nnlo3nf} shows the energy per
nucleon in symmetric nuclear matter for a wide range of densities
computed with MBPT2, CCD, and the CCD(T) approximation. The CCD
calculations included the 3NF in the normal-ordered two-body
approximation. The CCD(T) calculations were performed with 3NFs in the
normal-ordered two-body approximation (CCD(T+3NF$_{\mathrm{NO2b}}$)),
and going beyond the normal-ordered two-body approximation by
including the leading-order residual 3NF contribution to the
perturbative estimate for the $T_3$ amplitude
(CCD(T+3NF$_{\mathrm{NO3b}}$)). In these calculations we used $A=132$
nucleons with PBC and $n_{\mathrm{max}}=4$. For the densities we
consider here, the difference between PBC and TABC is small.

In contrast to calculations of neutron matter, the contribution from
the perturbative triples corrections is sizable in nuclear matter, and
about 1~MeV per nucleon in the range of densities shown when including
the 3NF in the normal-ordered two-body approximation.  Furthermore, we
find that the contribution of the residual 3NF to the CCD(T) energy is
significant around saturation density, indicating that the
normal-ordered two-body approximation for the 3NF might not be
sufficient in symmetric nuclear matter. We checked that the
contribution of the residual 3NF to the CCD amplitude equations is
negligible, and therefore it might be sufficient to include the full
3NF in the perturbative triples amplitude. In order to check the
accuracy of the perturbative triples approximation (CCD(T)) in nuclear
matter we also performed non-perturbative, iterative CCDT-1 (see
Refs.~\cite{lee1984a,lee1984b}) calculations for $A=28$ and
$n_{\mathrm{max}}=3$ at two different densities $k_F =
1.3~\mathrm{fm}^{-1}$ and $k_F = 1.6~\mathrm{fm}^{-1}$. We found that
the difference between CCD(T) and CCDT-1 in this range of densities is
at most 0.1~MeV per nucleon. Therefore, we conclude that the CCD(T)
approximation is accurate for the $NN$ potential NNLO$_{\mathrm{opt}}$
and chiral 3NFs in symmetric nuclear matter.

\subsection{Scheme dependence of three-nucleon forces}
In this Subsection, we try to further illuminate the role of 3NFs in
nucleonic matter. We study different regularization
schemes, and compute the energy per particle in pure neutron matter
and symmetric nuclear matter. The 3NF employed in the previous
Subsections exhibits a cutoff of $\Lambda=500$~MeV.  This cutoff is in
the momentum transfer, and therefore local in position
space~\cite{navratil2007}. This choice of regulator for the 3NF is
different from the regularization scheme that is used in the
nucleon-nucleon sector, and from other regularizations of the 3NF that
exhibit cutoffs on Jacobi momenta~\cite{epelbaum2002}. We note that
regulators that cut off initial and final Jacobi momenta lead to 
non-local interactions. Here, the cutoff function is
\[
f_R(\vec{p},\vec{q}) = \exp\left[
-\left( \frac{4p^2+3q^2}{4\Lambda^2}\right)^n\right] \ , 
\]
with $\vec{p}= (\vec{k}_1 - \vec{k}_2)/2$ and $q = [\vec{k}_3
-(\vec{k}_1 - \vec{k}_2)/2](2/3)$.  This regulator reduces to the
regulator used in the $NN$ sector for $\vec{q}=0$.  In the $NN$
potential NNLO$_{\rm opt}$ we use $n=3$, while for the local regulator
of the 3NF defined in Ref.~\cite{navratil2007} we use $n=2$ in the
exponential. In what follows, we compare the NNLO$_{\rm opt}$
interaction with a 3NF that also uses a local regulator but a lower
cutoff of $\Lambda=400$~MeV, and with a 3NF that employs a nonlocal
regulator and a cutoff $\Lambda=500$~MeV in relative Jacobi momenta.

\begin{figure}[hbt]
\includegraphics[width=0.9\columnwidth]{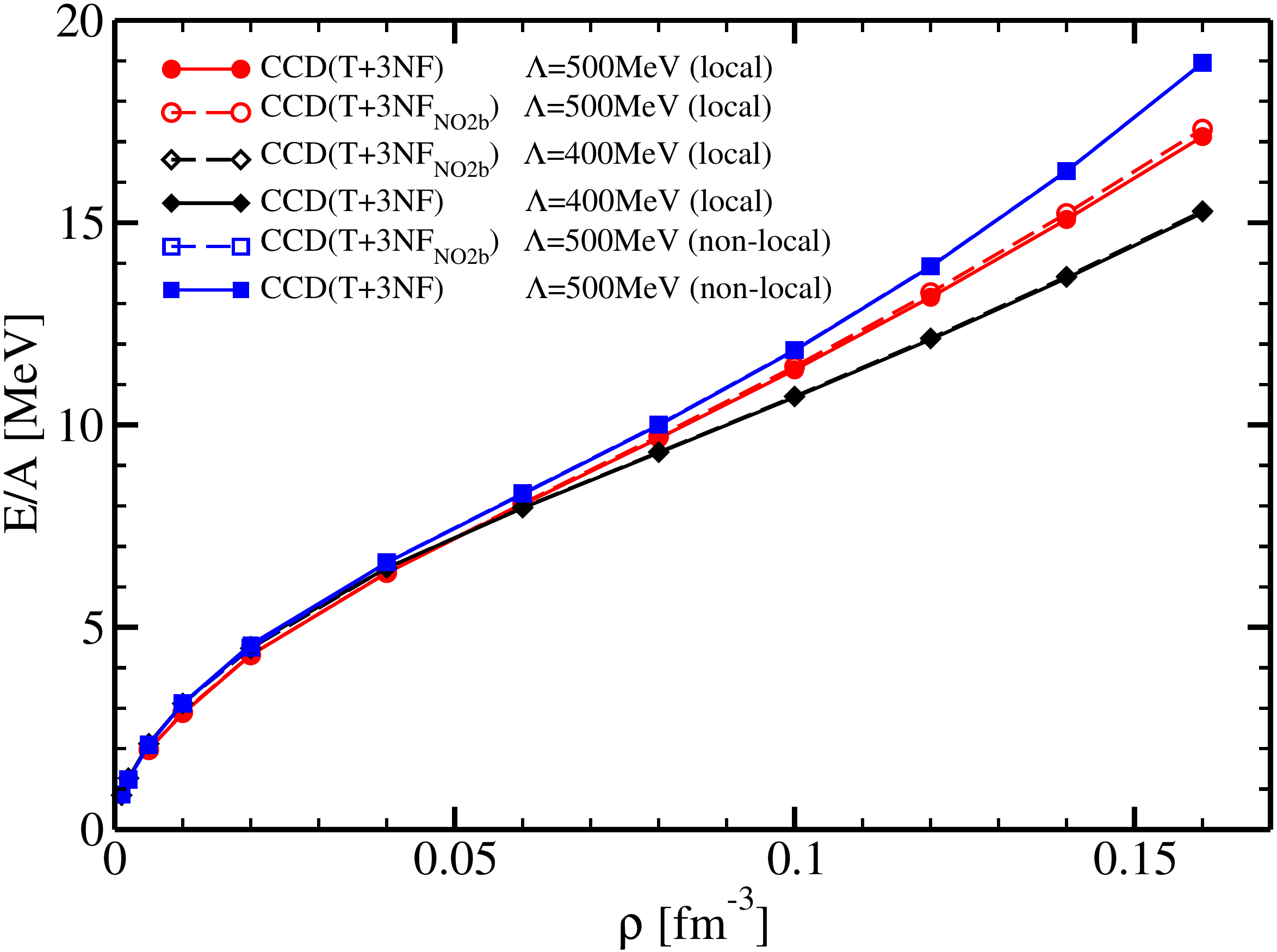}
\caption{(Color online) Energy per particle in pure neutron matter
  with NNLO$_\mathrm{opt}$ and 3NF computed in the CCD(T)
  approximation including 3NFs in the normal ordered two-body
  approximation and including the residual 3NF in the CCD(T$:wT_2=0$)
  approximation. For the 3NF we used a local regulator with cutoffs
  $\Lambda=400$ and $\Lambda =500$~MeV. The 3NF LECs are given by $c_E
  = -0.389$ and $c_D = -0.39$ for the $\Lambda =500$~MeV local
  regulator, while for the $\Lambda=400$~MeV local regulator we used
  $c_E =-0.27$ and $c_D=-0.39$ with $c_E$ adjusted to the $^4$He
  binding energy. For the non-local regulator with $\Lambda =500$~MeV
  cutoff we used $c_E =-0.791$ and $c_D=-2$ adjusted to the triton and
  $^3$He binding energies.  The calculations used 66 neutrons,
  $n_{\mathrm max}=4$, and PBC}
\label{fig:scheme2}
\end{figure}

Figure~\ref{fig:scheme2} shows the energy per particle in pure neutron
matter computed in the CCD(T) approximation. Here we included 3NFs
in the normal ordered two-body approximation, and in the
CCD(T$:wT_2=0$) approximation. For the latter, we went beyond the normal
ordered two-body approximation and included the residual three-body
term $w$ that enters at first order in the triples equation for $T_3$.
In neutron matter the contribution from the residual 3NF $w$ to the
energy per particle is small. This indicates that the normal-ordered
two-body approximation works very well. In the EoS calculation with
the local regulator and the lower cutoff $\Lambda=400$~MeV we adjusted
the LECs of the three-body contact term to $c_E=-0.27$ and kept $c_D$
unchanged. Then, the binding energies of the triton and the nuclei $^{3,4}$He  are
close to the experimental values. For the non-local regulator with cutoff
$\Lambda=500$ and power $n=2$ in the exponential, the LECs
$c_E=-0.791$ and $c_D=-2$ reproduce the triton and $^3$He binding
energies. In pure neutron matter the contributions from the 3NF
contact terms with the LECs $c_E$ and $c_D$ vanish for a non-local
regulator, and the contribution to the EoS depends only on the
pion-nucleon couplings $c_1$ and $c_3$ of the long-range two-pion
exchange term of the 3NF~\cite{hebeler2010b}. However, for a local
regulator the 3NF contact terms do not vanish in neutron matter
\cite{lovato2012}. The results for the EoS for pure
neutron matter show a regulator dependence at densities beyond $\rho =
0.08~\mathrm{fm}^{-3}$. The band obtained from the 
different 3NF regulators are within the corresponding band for neutron matter
obtained in Ref.~\cite{krueger2013}.

\begin{figure}[hbt]
\includegraphics[width=0.9\columnwidth]{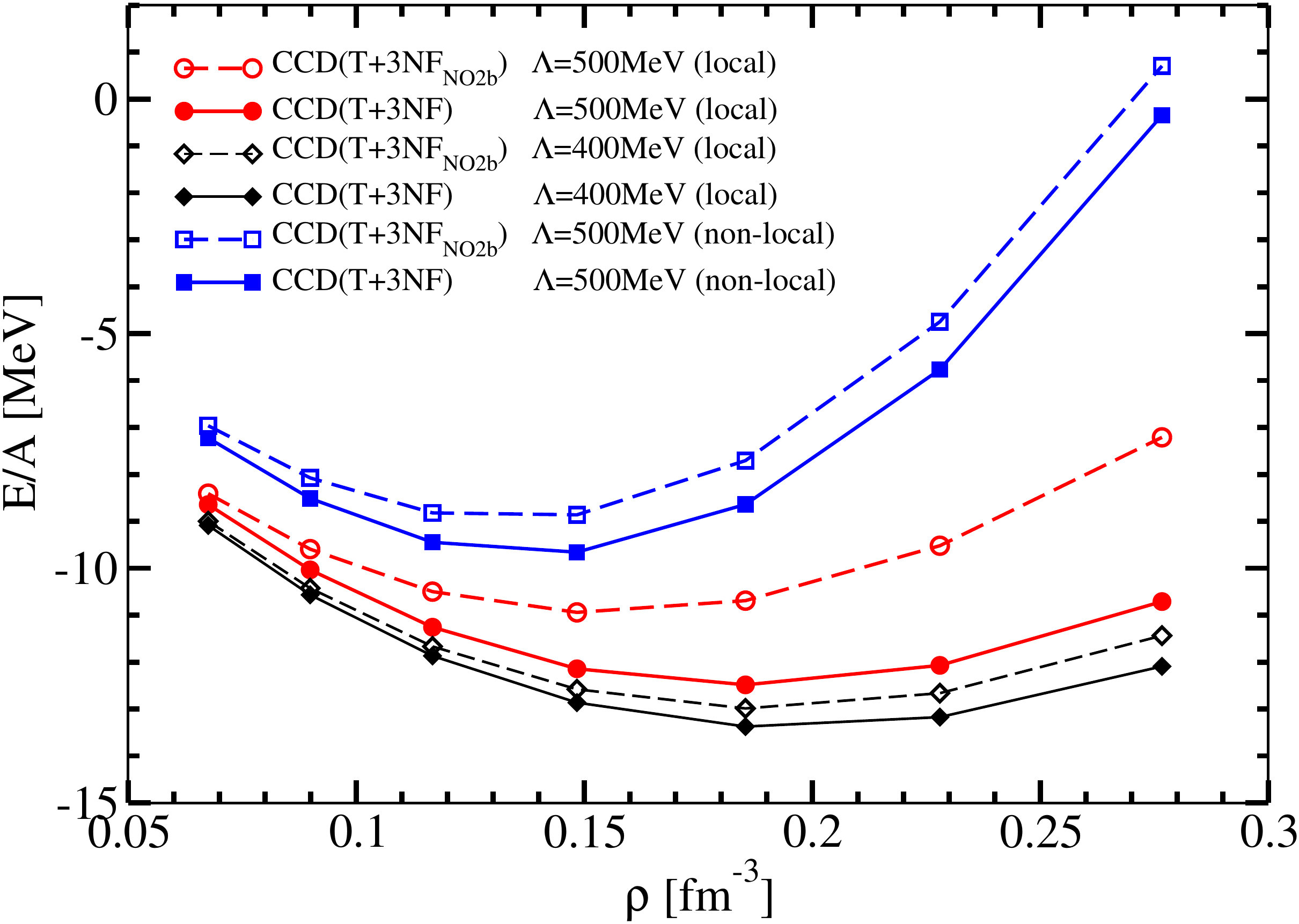}
\caption{(Color online) Same caption as in Fig.~\ref{fig:scheme2}, except that
  the energy per particle is for symmetric nuclear matter. The
  calculations used 132 nucleons, $n_{\mathrm max}=4$, and PBC}
\label{fig:scheme1}
\end{figure}

Figure~\ref{fig:scheme1} shows the corresponding plot for the energy
per particle in symmetric nuclear matter. Here the results for the
local regulator with a cutoff $\Lambda=500$~MeV exhibit a considerable
enhancement of the contribution from the residual 3NF $w$ to the
energy per particle at densities above the saturation densities. The
sizeable triples contribution of the residual 3NF $w$ questions the
usually observed hierarchy of the coupled-cluster approximation.  The
results from the lower cutoff $\Lambda=400$~MeV are much more
satisfactory in the sense that the contribution from the residual
three-body part $w$ to the binding energy per particle is considerably
smaller, and at the order of 0.5~MeV or less for the densities
considered. Likewise, the results obtained with the non-local
regulator at the cutoff $500$~MeV are also satisfactory in the sense
that the contribution from the residual 3NF $w$ is at most $1$~MeV to
the energy per particle at densities beyond the saturation point. One
might speculate whether this problematic feature of the local
regulator with a cutoff $500$~MeV is related to the large cutoff
dependence found in finite nuclei using this regulator
\cite{roth2012}. Naively one would expect that regulator dependencies
are higher-order corrections in an EFT. The large scheme dependencies
observed in Fig.~\ref{fig:scheme1} might therefore suggest that the
cutoff $\Lambda = 500$~MeV is too close to the EFT breakdown scale.

\begin{figure}[hbt]
\includegraphics[width=0.9\columnwidth]{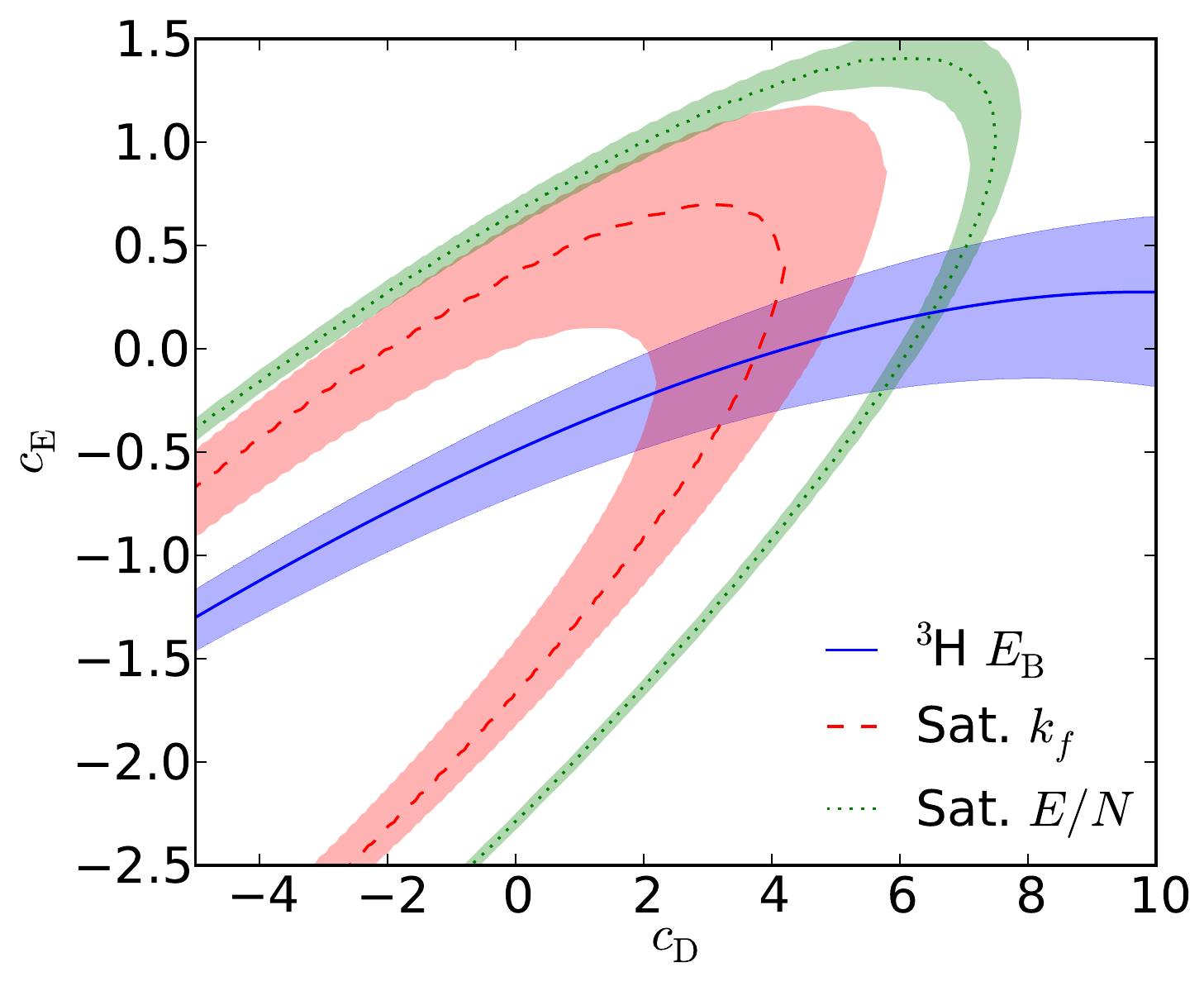}
\caption{(Color online) The blue band shows the region where the
  triton binding energy is reproduced within 5\% of the experimental
  value. The red band shows the region where the saturation fermi
  momentum in symmetric nuclear matter is reproduced within 5\% of its
  empirical value, and the green band shows the region where the
  energy per nucleon is within 5\% of the empirical value.}
\label{fig:cdce}
\end{figure}

For local and non-local regulators we considerably underbind nuclear
matter.  The saturation density for the local regulators is too high,
while for the non-local regulator the saturation density is closer to
the empirical value. We tried to adjust the LECs $c_E$ and $c_D$ such
that an acceptable result could be obtained simultaneously for the
saturation point in symmetric nuclear matter and the triton binding
energy. For the non-local regulator the result is shown in
Fig.~\ref{fig:cdce}. The blue band shows the region where the triton
binding energy is reproduced within 5\%. The red band shows the region
where the saturation Fermi momentum is within 5\% of its empirical
value, and the green band shows the region where the energy per
nucleon is within 5\% of the empirical value. The nuclear matter
calculations were obtained from MBPT2 calculations using 28 nucleons,
and we accounted for about 1~MeV per nucleon in missing correlations
energy, and about 0.5~MeV per nucleon due to finite size effects. It
thus seems that a simultaneous reproduction of saturation in light
nuclei and infinite matter is not possible without adjusting other
LECs.  As an example we considered the point $c_E = 0.3$ and
$c_D=-2.0$. This yields the saturation point $k_F\approx
1.4~\mathrm{fm}^{-1}$ and $E/A \approx 15.5$~MeV, while the triton
binding energy is $-13.53$~MeV.

\begin{figure}[thb]
\includegraphics[width=0.9\columnwidth]{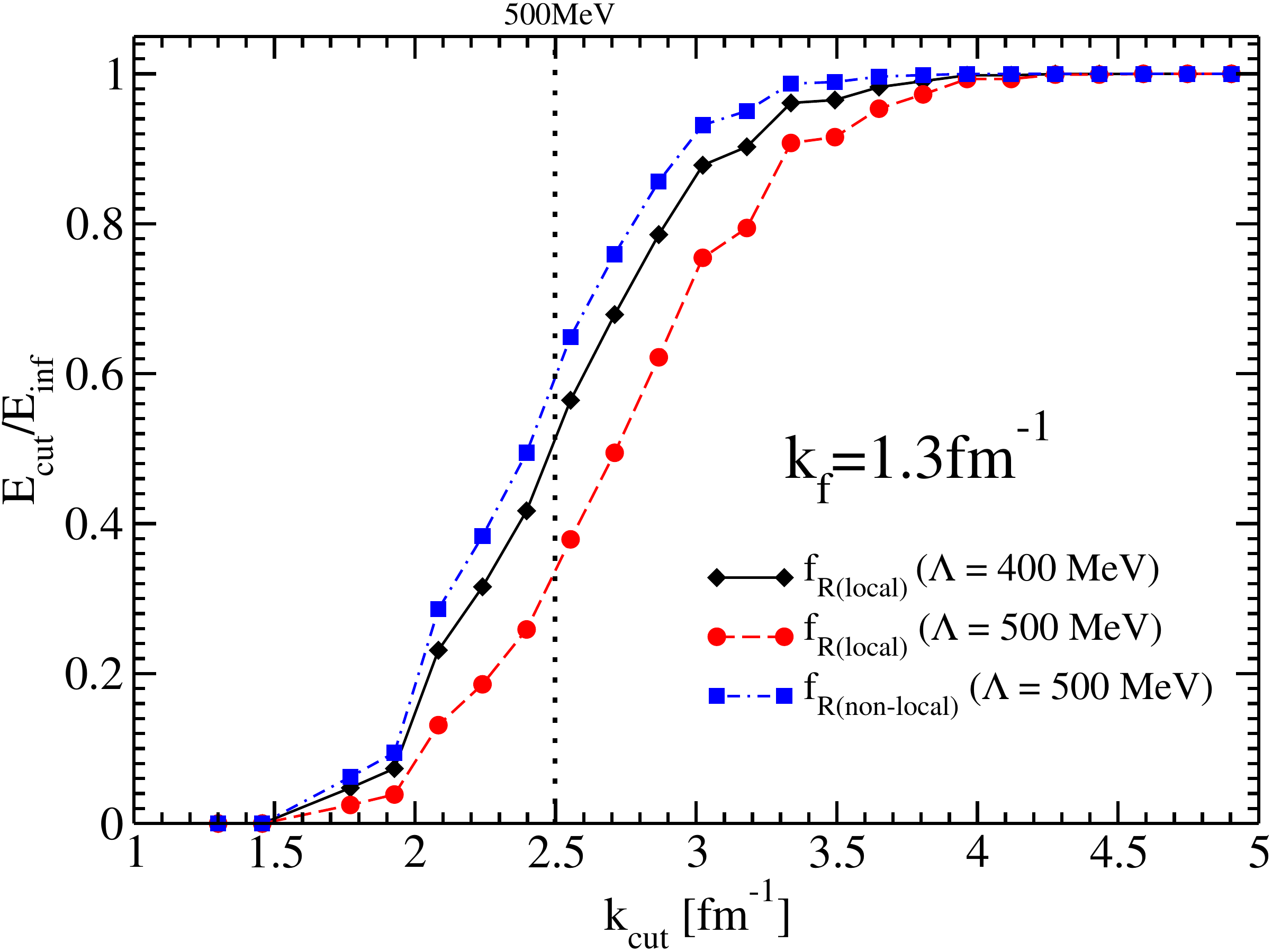}
\caption{(Color online) Cutoff dependent fraction of the residual 3NF contribution to 
the MBPT2 energy per particle in symmetric nuclear matter for
different regulators.  Results are shown for the local regulator with
$\Lambda = 500$ MeV (diamonds), the local regulator with $\Lambda =
400$ MeV (squares), and for the non-local regulator with $\Lambda=500$
MeV (circles).  Calculations used 132 nucleons, $n_{\mathrm max}=4$,
and PBC.}
\label{fig:scheme}
\end{figure}

We would like to understand better the role that different regulators
and cutoffs play for the chiral 3NF. Unfortunately, it is difficult to
visualize 3NFs in momentum space~\cite{hebeler2012,wendt2013}.  We
therefore compute the MBPT2 contribution of the residual 3NF $w$ and
cut off the involved momentum integrations at a single-particle
momentum $k_{\rm cut}$. Figure~\ref{fig:scheme} shows the fractional
contribution of the MBPT2 energy correction of the residual 3NF as a
function of $k_{\rm cut}$ at the Fermi momentum $k_F=1.3$~fm$^{-1}$.
The chiral cutoff of $\Lambda=500$~MeV is also shown as a dashed line
for comparison. We see that for the local cutoff $\Lambda=500$~MeV
most contributions to the MBPT2 result are from high single-particle
momenta that are well above the nominal chiral cutoff. The situation
is improved for the local regulator with lower cutoff
$\Lambda=400$~MeV and even more so for the nonlocal regulator with
cutoff $\Lambda=500$~MeV. For a discussion of different cutoff schemes
and convergence issues in calculations of the homogeneous electron gas
see Ref.~\cite{shepherd2012}.

Let us finally note that issues with 3NFs also arose in other
calculations. Lovato~{\it et al.}~\cite{lovato2012} pointed out that
the equivalence of different chiral 3NF contact
terms~\cite{epelbaum2002} is spoiled by local regulators. Roth~{\it et
  al.}~\cite{roth2011a} used SRG evolution to soften the chiral $NN$
interaction of Ref.~\cite{entem2003} combined with the local 3NF of
Ref.~\cite{navratil2007}, and found that the results in medium-mass
nuclei depend considerably on the SRG evolution scale. This dependence
is reduced for a cutoff $\Lambda=400$~MeV in the local
3NF~\cite{roth2012}. Clearly, more studies of chiral 3NFs are
necessary to fully understand regularization scheme dependences.

\section{Summary}\label{sec:summary}
We have performed coupled-cluster calculations of nucleonic matter with
interactions from chiral EFT at NNLO. The single-particle states
consist of a discrete lattice in momentum space, and the
implementation of twist-averaged boundary conditions mitigates shell
oscillations and finite-size effects. Our benchmark calculations agree
well with other well-established methods. We find that neutron matter
is perturbative, while symmetric nuclear matter is not perturbative,
with significant contributions beyond perturbation theory and particle
ladders.

For the employed $NN$ potential NNLO$_{\rm opt}$ and 3NFs, the neutron
matter results fall within the error estimates of previous
calculations for chiral interactions, with 3NFs acting
repulsively. For nuclear matter, the empirical saturation could not be
reproduced, and the results are very sensitive to the employed
regulator (local vs. nonlocal) and cutoff. At larger chiral cutoffs,
the nonlocal regulator is preferred over the local one because it
corresponds closer to the cutoff generated by the finite
single-particle basis. It seems that the variation of the 3NF contact
terms alone is insufficient to achieve both an acceptable saturation
point of nuclear matter and an acceptable binding of light nuclei.

\begin{acknowledgments}
  We thank S.~K.~Bogner, E.~Epelbaum, R.~J.~Furnstahl, A.~Mukherjee,
  and F.~Pederiva for discussions. This work was supported by the
  Office of Nuclear Physics, U.S. Department of Energy (Oak Ridge
  National Laboratory), under DE-FG02-96ER40963 (University of
  Tennessee), DE-FG02-87ER40365 (Indiana University), DE-SC0008499 and
  DE-SC0008808 (NUCLEI SciDAC collaboration), the Field Work Proposal
  ERKBP57 at Oak Ridge National Laboratory, the LDRD program at Los
  Alamos National Laboratory, and the Research Council of Norway under
  contract ISP-Fysikk/216699.  Computer time was provided by the
  Innovative and Novel Computational Impact on Theory and Experiment
  (INCITE) program. This research used resources of the Oak Ridge
  Leadership Computing Facility located in the Oak Ridge National
  Laboratory, which is supported by the Office of Science of the
  Department of Energy under Contract No. DE-AC05-00OR22725, and used
  computational resources of the National Center for Computational
  Sciences, the National Institute for Computational Sciences, and the
  Notur project in Norway. Computing time has also been provided by
  Los Alamos Open Supercomputing. This research also used resources of
  the National Energy Research Scientific Computing Center, which is
  supported by the Office of Science of the U.S. Department of Energy
  under Contract No. DE-AC02-05CH11231.
\end{acknowledgments}

\bibliography{refs}
\bibliographystyle{apsrev}

\end{document}